\documentclass{aastex62}

\newcommand{\ebtel}{{\tt EBTEL}}

\graphicspath{{./}{figures/}}
\received{27th December 2020}
\revised{13th May 2021}
\accepted{18th May 2021}
\submitjournal{ApJ}

\shorttitle{Hydrodynamics of small transient coronal brightenings}
\shortauthors{Rajhans, Tripathi, and Kashyap}

\begin{document}

\title{Hydrodynamics of small transient brightenings in Solar corona}

\correspondingauthor{Abhishek Rajhans}
\email{abhishek@iucaa.in}
\author[0000-0001-5992-7060]{Abhishek Rajhans}

\author[0000-0003-1689-6254]{Durgesh Tripathi}
\affiliation{Inter-University Centre for Astronomy and Astrophysics, Post Bag - 4, Ganeshkhind, Pune 411007, India}
\author[0000-0002-3869-7996]{Vinay L. Kashyap}
\affiliation{Center for Astrophysics $|$ Harvard \& Smithsonian, 60 Garden St.\ Cambridge MA 02138, USA }

\begin{abstract}

Small scale transients occur in the Solar corona at much higher frequencies than flares and play a significant role in coronal dynamics. Here we study three well-identified transients discovered by Hi-C and also detected by the EUV channels of Atmospheric Imaging Assembly (AIA) on board Solar Dynamics Observatory (SDO). We use 0-D enthalpy-based hydrodynamical simulations and produce synthetic light curves to compare with AIA observations. We have modeled these transients as loops of $\sim$~1.0~Mm length depositing energies $\sim$ $10^{23}$ ergs in $\sim$ 50 seconds. The simulated synthetic light curves show reasonable agreement with the observed light curves. During the initial phase, conduction flux from the corona dominates over the radiation, like impulsive flaring events. Our results further show that the time-integrated net enthalpy flux is positive, hence into the corona. The fact that we can model the observed light curves of these transients reasonably well by using the same physics as those for nanoflares, microflares, and large flares, suggests that these transients may have a common origin.
\end{abstract}

\section{Introduction} \label{sec:intro}
The presence of high-temperature ($>$1~MK) plasma in the solar corona was discovered in the 1940s. How this plasma, above the much cooler photosphere, is heated to such high temperatures has been one of the most challenging questions in astrophysics. Though our understanding of the energy dissipation in the corona has improved substantially, the full solution to the problem of coronal heating remains elusive and the transfer of mass and energy between layers of the solar atmosphere is not completely understood \citep[see e.g.,][for a review]{Klimchuk2006, reale}.
 Multi-wavelength observations of the Sun show that different layers couple through magnetic fields. By and large, theories related to omnipresent coronal heating fall into two groups: AC heating and DC heating \citep[see e.g.,][]{klim2015, Walire}. Depending on the frequency of occurrence, heating events can be classified into high and low frequency heating, with the former may contribute to steady and latter to transient events \citep[see e.g.,][]{TriKM_2011, amywinebarger}.

Solar flares provide the best observed examples of impulsive events taking place in the Solar atmosphere. \cite{Hudson_1991} conjectured that for maintaining the corona at a temperature greater than 1~MK with the help of impulsive events, there must be a large frequency of such events with smaller energy, and found that the relationship between number of events and their energies obeys a power law distribution $\frac{dN}{dE} \propto E^{-\alpha}$,  where $dN$ is the rate of occurrence of  events having energy in the range [$E,E+dE$] and $\alpha$ is a positive number \citep[see e.g.,][for a review]{HanHB_2011}. The power law distribution of flares of different  energies is a sign of underlying phenomenon of self organized criticality \citep{lu}. For the heating to be dominated by nanoflares, the  the power law index
$(\alpha)$ should be greater than 2. This has led to many observational studies \citep[see e.g.,][]{Shimizu_1995, BerCM_1998, BerMC_2001, KruB_1998, KruB_2000, ParJ_2000, AscNTW_2000, AscTNS_2000, ChrH_2008, HanCK_2008} based on the counting of different types of transient events which result in varied negative slopes of power law
ranging from $1.6\lesssim\alpha\lesssim2.2$. However, there are limitations to such studies due to constraints on the cadence, passbands, and resolutions of instruments. 
Also there remains a chance that flares of different energy, particularly at lower energies, are under counted \cite[see e.g.,][]{PauS_2007, UpeT}. Additionally, it is possible that different events may be generated due to different mechanisms and hence they would not necessarily follow the same power law distribution.

\cite{habbal} studied coronal bright points and used Ly-$\alpha$ emission as a proxy for conduction losses from the corona into the chromosphere. \cite{pres1999} were then able to establish that conduction losses are at least an order of magnitude larger than radiation losses, implying that radiation loss from the corona is a small fraction of total energy dissipated. The smallest brightenings detected thus far \citep{regnier, srividya} are due to the observations recorded by Hi-C \citep{ken}. \cite{ srividya} identified 27 such events in Hi-C images and
performed a detailed study to understand their energetics using simultaneous observations obtained with Atmospheric Imaging Assembly \citep[AIA;][]{aia} on board the Solar Dynamics Observatory (SDO). The study found conduction to be the dominant cooling mechanism in corona. This is a feature shared by impulsive events like flares, microflares, and nanoflares, suggesting that the same physical mechanism is shared by these small transient brightenings.  We note, however, that a number of simplifying assumptions, such as detailed thermal balance and stationary loop structures, were made in this study.

In this work we carry out hydrodynamic simulations to gain a theoretical understanding of the energetics of small brightenings identified by \citep{regnier, srividya}. We have selected three of the brightenings (BR-00, 07, 26) from \cite{srividya} for a detailed study, as they show the simplest profiles, with a single peak, and a clearly visible decay phase (see \S\ref{subsec:br00}, \S\ref{subsec:br07} and \S\ref{subsec:br26}). We have used a 0-D numerical code called Enthalpy Based Thermal Evolution of Loops  \citep[\ebtel;][]{Klimchuk2008,cargill}. The results obtained from {\ebtel} simulations were used along with AIA response functions to produce synthetic lightcurves, mimicking observations \citep{srividya}. The rest of the paper is arranged as follows. In \S\ref{sec:obs} we outline the observational results of \cite{srividya} including the assumptions made. In \S\ref{mod} we briefly discuss {\ebtel}, the relevant input parameters, and the simulation set up. The analysis is described in \S\ref{sec:analysis} and results are discussed in \S\ref{sec:results}. We summarise our results in \S\ref{sec:cad}.

\section{ \textbf{Data}} \label{sec:obs}
 
\begin{figure}[hb!]
\centering
\includegraphics[width=\linewidth]{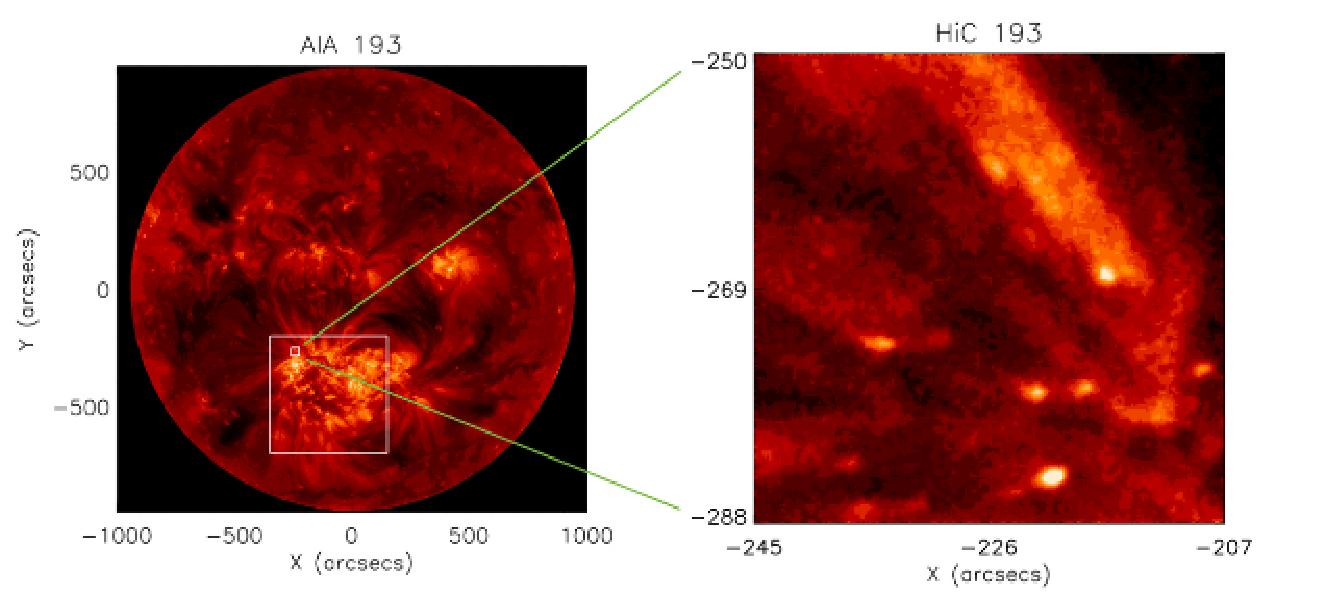}	
\caption{The site of the bright points considered here. {\sl (Left)} The full disk AIA/193\,\AA~filter image of the Sun, with an inset box showing the Hi-C field of view (large square box), as well as the location of the fan-loop structure where bright points were detected (small square box).  {\sl (Right)}  The Hi-C image corresponding to the small square box showing the location of the fan loop region where the transient brightenings were observed.
The figure is taken from \cite{srividya}.} \label{fig:sri}
\end{figure}

Hi-C (High-Resolution Coronal Imager) is a sounding rocket mission that observed the Sun in the 193\,\AA\ passband filter with a pixel size of 0.1~$\arcsec$ \citep{ken}. It was launched on July 11, 2012 and recorded observations of active region AR\,11520 for $\approx$5 minutes. One of several interesting phenomena observed in detail\footnote{https://hic.msfc.nasa.gov/publications.html\#hic1\_{pubs}} were multiple tiny brightenings within a system of fan loops rooted in the active region \citep{stephane, srividya}.

\cite{srividya} identified 27 such point-like brightenings using the automatic detection algorithm of \cite{SubMD_2010}, and performed a detailed study of energetics involved in these events. For this purpose, the Hi-C observations were cross calibrated with simultaneously obtained SDO/AIA (Solar Dynamics Observatory/Atmospheric Imaging Assembly; \cite{aia,OdwDMWT_2010}) data.

AIA provides full disk images of the solar atmosphere in 7 separate EUV passbands with a cadence of 12 s and pixel size of 0.6\arcsec. Fig.~\ref{fig:sri} displays the AIA full disk image recorded in the 193~{\AA} bandpass filter, the Hi-C field of view (larger inset white box) and the region where brightenings were found (smaller inset white box). The latter is zoomed-in in the right panel of Fig.~\ref{fig:sri}.

In order to determine the thermal structure of the brightenings \cite{srividya} obtained the \textit{Differential Emission Measure} (DEM) using six optically thin filters of AIA. For this purpose the \texttt{PINTofALE} package \citep{KasD_1993} was used. The DEMs were used for computing emission measure weighted temperatures and electron densities. It was found that all these events have temperatures $\approx 10^{6.2-6.6}$ K, electron number densities  
of the order of 10$^{9}$~cm$^{-3}$  
and estimated radiative energy losses of 10$^{24-25}$~ergs.  

Assuming that a magnetic flux density of $\sim$50~G is involved in each of these brightenings, \cite{srividya} estimated the total magnetic energy to be  
of the order of $10^{26}$ ergs and hence should be sufficient to power these brightenings with radiative energies of order of 10$^{24}$ ergs. 
On the basis of cooling timescales in static equilibrium, they suggested  
that thermal conduction was the dominant cooling mechanism in corona for these brightenings.  

It is important to note that the estimates of \cite{srividya} are crude, as they rely on stationary equilibrium estimates of conduction and radiative cooling timescales of corona and are integrated estimates over the lifetimes of the brightenings. Furthermore these neglect flows and density variations. Hence it is necessary to, first, validate the results using more realistic dynamical loop constructs, and second, to identify the region of the parameter space, where loops that can mimic the observed light curves exist. 
 
\section{Hydrodynamic modelling} \label{mod}

In this work, we have performed 0-D simulations using single fluid \ebtel~\citep{Klimchuk2008, cargill}.
The idea behind 0-D description is to study the time evolution of length averaged quantities. It is based on 1-D field aligned simulations showing that temperature, density and pressure are within factors of a few between the base and apex of coronal part of loop in 1-D simulations. As a result, the corresponding length averaged quantities are characteristic of coronal values.  The reduced computational cost and improved speed of computing 0-D solutions makes studying the temporal evolution of length-averaged physical quantities a fruitful alternative to carrying out detailed numerical hydrodynamical simulations. It was demonstrated that the results obtained with the \ebtel~differ from 1-D hydrodynamic simulations by at most 15-20\% despite being much faster (few seconds on a contemporary laptop).

\subsection{Enthalpy Based Thermal Evolution of Loops (\ebtel)}\label{ebtel}
 
\begin{figure*}[h!]
\centering
\includegraphics[width=\linewidth]{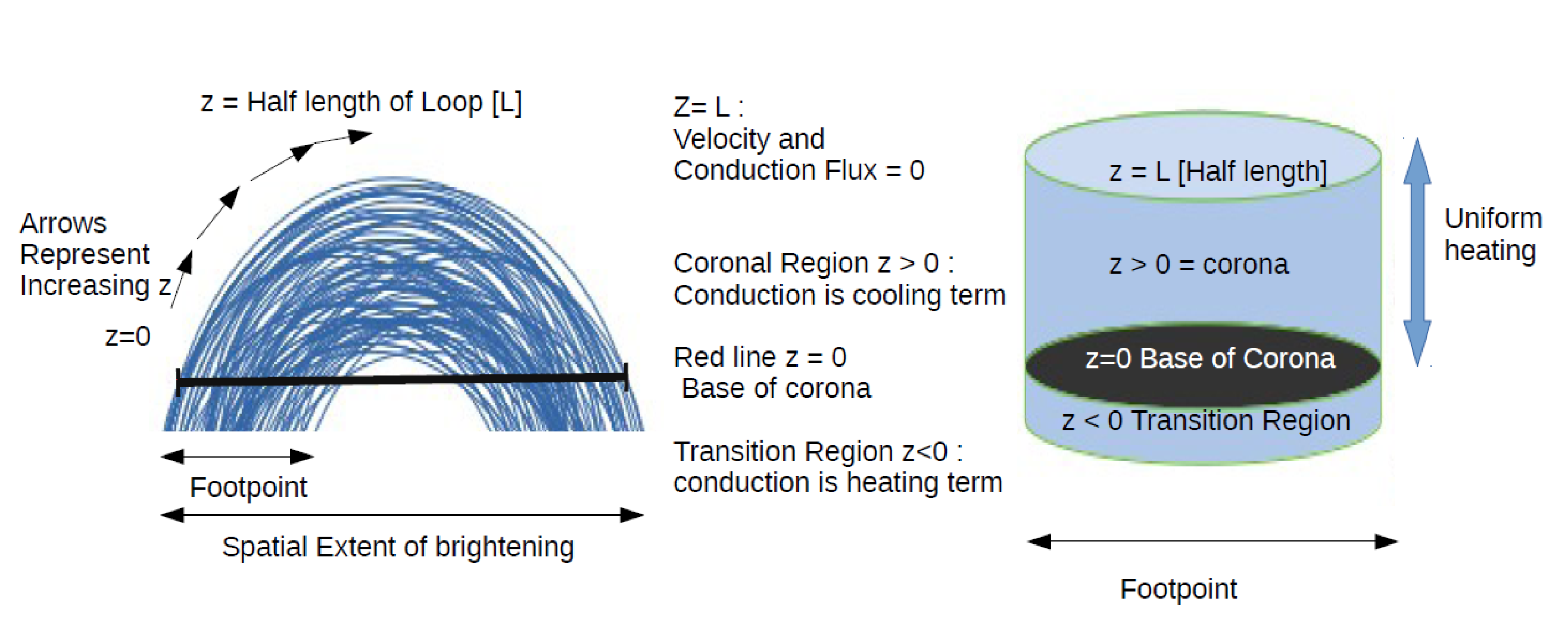}	
\caption{Sketch of the loop set-up. {\sl (Left)} A flux tube filled with multiple strands, rising up from the chromosphere at the bottom through a loop footpoint, and submerging at another footpoint.  The thick horizontal line represents the base of the corona, and the loop is assumed to be symmetrical around the apex.  {\sl (Right)} The representation of half of the loop in \ebtel.  The plasma aligned along the field lines are shown straightened out, with the various regions corresponding to the physical system labeled. } \label{fig:cartoon}
\end{figure*}

\ebtel~computes length-averaged quantities over time as they respond to varying heat input. It relies on a simplified form of the Navier-Stokes equation where flows are assumed to be subsonic and hence the kinetic energy term is neglected. Heating is uniform both along and across the loop. The loop is assumed to have a uniform cross section \citep{klimcrosec}, and symmetric about the apex. Due to this symmetry, vector quantities like conduction flux and velocity vanish at the loop top.
The base of corona is defined such that conduction is a cooling term in corona and heating term in transition region. Across the base of transition region, any conduction or enthalpy flux is assumed to be negligible. Fig.~\ref{fig:cartoon} depicts the sketch of loop setup along with the different regions and properties of the loop. 

\subsection{Input parameters} \label{subsec:inputpar}

\begin{figure*}[h!]
\centering
\includegraphics[width=0.5\textwidth]{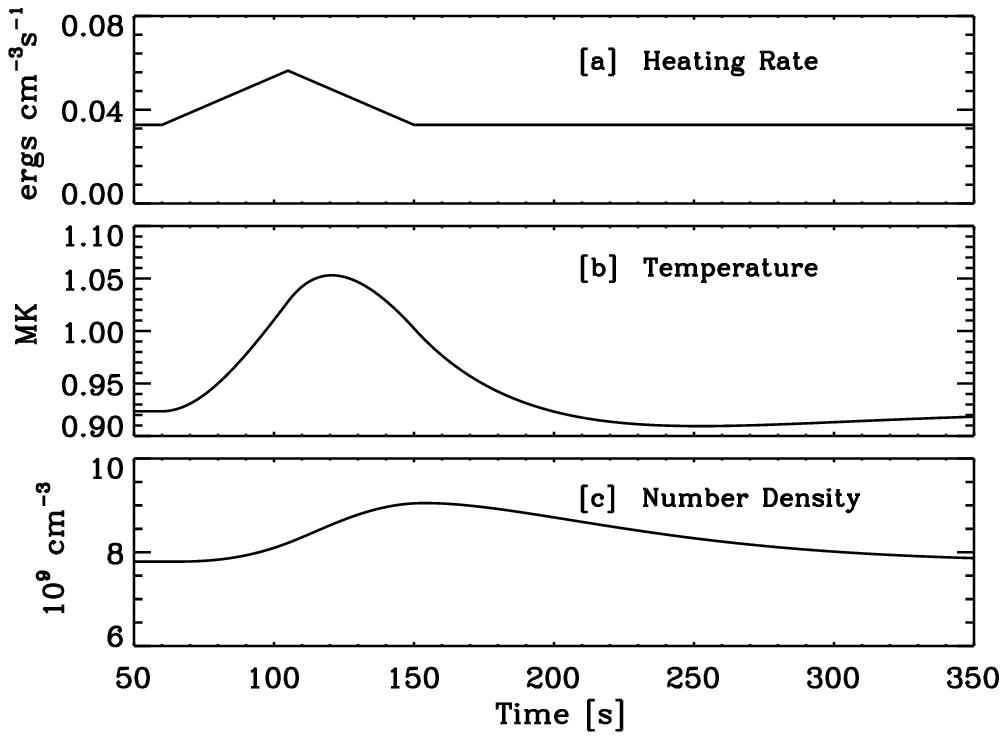} 
\caption{Response of a loop of half length of 1.0~Mm to a triangular heating function computed using \ebtel. (a) The heat given per unit volume to the loop in units of ergs~cm$^{-3}$~s$^{-1}$. (b) Time evolution of temperature in MK (c) Time evolution of number density in 10$^{9}$ cm$^{-3}$ }\label{fig:generic}
\end{figure*}

In order to run the \ebtel\ simulations, we have to specify some fundamental input parameters. \ebtel\ requires two inputs: the half length of the loop and the heating function. The  heating  function is defined as the rate at which the heat is  deposited in the loop per unit volume (in units of ergs~cm$^{-3}$~s$^{-1}$). An illustrative example of how \ebtel\ is used, is shown in Fig.~\ref{fig:generic}. The upper panel shows the heating function, applied to a loop of half length 1.0~Mm. The energy deposited begins to rise linearly (starting at 60~s) from an ambient value to a peak (at 105~s) and then declines linearly back to the ambient value (at 150~s).  The baseline ambient level is necessary to establish the presence of a corona, in this case at a temperature of 0.92~MK and a density of 7.8$\times 10^{9}$~cm$^{-3}$.  This triangular heating pulse leads to an increase in the temperature of the plasma in the loop (middle panel), which reaches a maximum at time close to the heating.  In contrast, the plasma density (bottom panel) rises more shallowly, reaches a peak after the plasma fills the loop in response to the enthalpy flux from the transition region before it declines.  Notice that the loop temperature drops below the ambient value after the heating pulse is ended. This is due to the heated plasma becoming over dense and cooling rapidly (due to $T \propto n^2$ scaling law ) until the excess plasma is drained.


\subsection{Synthetic lightcurves}\label{subsec:fm}
In order to compare observations with theoretical predictions, we need to compute synthetic intensities using simulations. For a given instrument's filter response $R(T)$ as a function of temperature $T$, the expected intensity, 
\begin{equation}\label{eq:intensity}
I = \int DEM(T) \cdot R(T)~dT ~~{\rm [DN~s^{-1}~pix^{-1}]} \,,   
\end{equation}
where $DEM(T)$ is the differential emission measure\footnote{The units of $R(T)$ and $DEM(T)$ in the solar context are typically [DN~cm$^{5}$~s$^{-1}$~pix$^{-1}$] and [cm$^{-5}$~K$^{-1}$] respectively.} over the temperature range $[T,T+dT]$.  In \ebtel, it includes the contributions of both the corona and the transition region.

\subsection{Estimation of Conduction and Radiation Losses and Enthalpy} \label{subsec:condrad}

For studying the various cooling processes and relative importance of different energy terms, we need to compute the energy fluxes associated with conduction, radiation and enthalpy from outputs provided by \ebtel.  

The conduction flux \citep{Klimchuk2008} 
\begin{equation}\label{eq:condloss}
F_{c} = \frac{F_{sp}F_{sat}}{\sqrt{F_{sp}^{2}+F_{sat}^{2}}} \,,
\end{equation} 
with
$F_{sp}$ and $F_{sat}$ being the Spitzer and saturation fluxes respectively and 
\begin{equation}\label{eq:spitzer}
F_{sp}=-\frac{2}{7}\kappa_{0}\frac{T_{a}^{7/2}}{L} \\
\hspace{0.3cm} \& \hspace{0.3cm} 
F_{sat}=-0.25\sqrt{\frac{k_{b}^{3}}{m_{e}}}\bar{n}\bar{T}^{1.5} \,,
\end{equation}
where $T_{a}$ is the temperature at the apex of loop, 
\textbf{$\kappa_{0}=8.12\times10^{-7}$ [cgs]}, $k_{b}$ is Boltzmann's constant, $m_{e}$ is the electron mass, and $\bar{n}$ and $\bar{T}$ are length-averaged electron number density and temperature. 
The radiation flux, 
\begin{equation}\label{eq:radloss}
F_{R} = (1+c_{1})\bar{n}^{2}L\Lambda(\bar{T}) \,,
\end{equation}
where $c_{1}$ is the ratio of the total radiation losses from the transition region and the corona as computed within \ebtel, and $\Lambda(\bar{T})$ is the optically thin power loss function of temperature, and is computed using the routine 
{\tt rad\_loss.pro} from Solarsoft software, using the {\sl Chianti} database \citep{dere1997,landi2013} and coronal abundances from \cite{grevesse2007}.

The enthalpy flux through the base of the loop footpoint ($z=0$ in Fig.~\ref{fig:cartoon}),
\begin{equation}\label{eq:enthalpy}
F_{P} = \frac{5}{2}\bar{P}v_{0} \,,
\end{equation}
where $\bar{P}$ is the average pressure in the coronal part of the loop and $v_{0}$ is the velocity at the base of the loop.

\section{Generation of Synthetic Light Curves}\label{sec:analysis}

We probe the range of input parameters for generating synthetic light curves consistent with observations. For this purpose, we take the observations recorded in the AIA\,193~{\AA} filter as reference. This is primarily because the lifetimes of observed brightenings were obtained using this filter by \cite{srividya}. 
In order to determine the input parameters that best describe the transients, we employ two methods: in the first, we assume that the transients are part of the same dynamical system as the ambient corona, and use the pre-transient intensity in AIA\,193 to constrain the inputs (\S\ref{subsec:method1}); in the second, we assume that the transients and the ambient corona are dynamically distinct, we match the characteristics of the background-subtracted light curves (\S\ref{subsec:method2}).

\subsection{Method 1: Modeling the transients with background}\label{subsec:method1}

Here we seek to find
a suitable combination of 
heat input
and loop half length 
that 
replicates the observed
background intensities, rise time of the transient, and the average background subtracted intensities of the transient as observed in AIA\,193\,{\AA}.

\subsubsection{Range of parameters used in simulations}\label{subsubsec:analysism1p1}
\cite{srividya} estimated the spatial extent of the bright points to be $\approx$ 2$\times$2 AIA pixels. Assuming that the loop would be semicircular, 
the expected half length is
$\approx$ 0.65~Mm. 
In order to account for
uncertainties, we simulate loops with half lengths ranging between 0.1{--}1.5~Mm. The heating function provided in {\ebtel} 
consists of two parts, viz.\ steady heating that creates the background and time dependent heating that causes the transient. For each loop in this range, we set up a background heating ($Q_{bkg}$) in accordance with the scaling law used in {\ebtel}, 

\begin{equation}\label{eq:background}
Q_{bkg}~~{\rm [ergs~cm^{-3}~s^{-1}]}  ~~ \approx ~~ \frac{2}{7}\kappa_{0}\frac{T_{a}^{\frac{7}{2}}}{L^{2}} =\frac{2}{7}\left(\frac{10}{9}\right)^{\frac{7}{2}}\kappa_{0}\frac{\bar{T}^{\frac{7}{2}}}{L^{2}} 
\end{equation}
where $\kappa_{0}=8.12\times 10^{-7}$ in cgs units, $T_{a}$ is the temperature at loop summit, $\bar{T}$ is average temperature of the coronal part of loop and $\bar{T}\approx{0.9}T_{a}$ \citep[][]{cargill}. We seek solutions where $\bar{T}\approx1$~MK. 

To simulate the transient, we use a triangular heating profile characterized by a maximum heating rate $H_{m}$ ergs~s$^{-1}$~cm$^{-3}$, and a total duration $t_{dur}$ seconds \citep[see e.g.,][]{cargill, Klimchuk2008}. Then the total energy $E$ dissipated in the semi-loop,

\begin{equation}\label{eq:peakht}
E = \frac{1}{2}~A_{\rm foot}~L~t_{dur}~H_{m} ~~ {\rm [ergs]} \,.
\end{equation}

For the exemplar case shown in Fig.~\ref{fig:generic}, we have $H_{m}=0.2$~ergs~s$^{-1}$~cm$^{-3}$ and $t_{dur}=90$~s with cross-sectional area $A_{\rm foot}=1.76\times 10^{15}$~cm$^2$, corresponding to one AIA pixel. Note that to fix the time dependent heating function, we vary the total energy budget instead of volumetric heating rate. We run the simulations by varying the amount of total energy in the range $10^{22}$ to $10^{25}$~ergs, in steps of 0.1~dex. This range was chosen as it brackets the radiative losses for these transients as determined by \cite{srividya}. The typical life time of these transients of the order $\approx100$~s and therefore we consider $t_{dur}$ ranging from 10 to 200~s in steps of 10~s.

\subsubsection{Parameters for specific brightening}\label{subsubsec:analysism1p2}

Our goal is to identify the set of parameter values (viz., loop half length, heating duration, and heating rate, $[L,t_{dur},E]$) which mimic the characteristics of the observed brightenings. To find the best parameter set, we seek to compare and match the following characteristics of the observed and simulated light curves:

\begin{enumerate} 
\item The average background level of observed light curves and the background subtracted intensities averaged over the lifetime of the events as observed in AIA\,193\,{\AA} images
\item The rise times of the events in the observed light curves obtained from AIA\,193\,{\AA} filters. We require to match the rise times instead of total duration because the decay times are subject to large systematic errors due to the difficulty of identifying precisely when the model intensities become indistinguishable from the background.
\end{enumerate}

We find that for all the three brightenings considered in this study, there exist physically plausible loop lengths, heating rates, and duration that capture the behavior of the observed AIA light curves. We emphasize, however, that the best parameters sets were not obtained by performing fits to the data; such a process would be unrealistic given the simplicity of the models we consider. To identify the input parameters for {\ebtel}, we follow the following procedure:

\begin{enumerate}
\item First, we identify the most suitable loop half length. We note from Eq.~\ref{eq:background} that for a given temperature, the background heating is a function of loop length. Therefore, we compare the background intensities in the light curves of AIA 193~{\AA} with those obtained using Eq.~\ref{eq:background} for all values of $L$ within the range 0.1{--}1.5~Mm (see \S\ref{subsubsec:analysism1p1}), while fixing the temperature at 1~MK. We select that value of $L$ that provides the background intensity closest to the observed values in the 193~{\AA} light curves.
\item Second, we identify the total duration of the heating events. For this, we deposit an heating event with total energy $E = 10^{22}$~ergs by varying the time duration $t_{dur}$ within the range of 10 to 200~s (see \S\ref{subsubsec:analysism1p1}), for the values of $L$ previously obtained. Given $(L,E)$, the rise times are primarily dependent on $t_{dur}$. The values of $t_{dur}$, which give the closest agreement with the rise times of the observed events are selected.
\item Finally, for the given $L$ and $t_{dur}$, the average intensity of the brightening in the 193~{\AA} filter is estimated for values of $E$ ranging from $10^{22}-10^{25}$~ergs. The value of $E$ which 
generates an intensity closest to that observed is then selected.
\end{enumerate}

\subsection{Method 2: Modeling the transients without background}\label{subsec:method2}

In contrast to the method outlined above, here we detail an alternate method where background levels are ignored and only the rise time and average intensities of the transients are matched with model predictions to obtain a suitable set of \ebtel\ parameters.  This effectively treats the transients as dynamically distinct events in relation to the ambient corona.

\subsubsection{Parameter range used in simulations}\label{subsubsec:analysism2p1}
Using the assumption of semicircular loops confined within an area equivalent to 2$\times$2 AIA pixels, we fix the loop half length to be 0.65~Mm. The background heating rate is generally taken to be two to three orders of magnitude smaller than the main heating event \citep[][]{Klimchuk2008, cargill}. For a loop length of order of 1~Mm with cross-sectional area of 1 AIA pixel, we require a volumetric heating rate of the order of 0.01 to 0.1~ergs~cm$^{-3}$~s$^{-1}$, such that $\sim 10^{23}$ ergs is deposited in $\sim 50 $ s. Hence we have set a uniform background heating rate of $10^{-4}$~ergs~cm$^{-3}$~s$^{-1}$, such that the background temperature and densities are an order of magnitude lower than the peak values. 

The remaining task is to find the heating duration and total energy budget for the event. They were determined in the same way as mentioned in \S\ref{subsubsec:analysism1p1} once loop length and background heating rates 
are set 
After fixing the half length of the loop and background heating rate to 0.65~Mm and $10^{-4}$~ergs~cm$^{-3}$~s$^{-1}$, respectively, we follow steps 2 \& 3 enumerated in \S\ref{subsubsec:analysism2p1} to obtain the doublet $[t_{dur},E]$ (see Table~\ref{table:inputparam}).

\section{Results}\label{sec:results}
Using the two methods described above, we identify the triplet $[L,t_{dur},E]$ of inputs required for {\ebtel} that best describes the transient under under study or under investigation (see Table~\ref{table:inputparam}). We plot the observed and simulated light curves for AIA\,193\,{\AA} (panels a \& d), simulated plasma temperature (panels b \& e) and density (panels c \& f) obtained from both methods; explicitly modeling background levels (left column: method 1; \S\ref{subsec:method1}) and excluding background from the modeling (right column: method 2; \S\ref{subsec:method2}) for BR00, BR07 and BR26 in Figs.~\ref{fig:br00res}, \ref{fig:br07res} and \ref{fig:br26res}, respectively. We have investigated the robustness of both models by bracketing the nominal duration by half and twice the selected $t_{dur}$. Moreover, we have also obtained the synthetic light curves using both methods for other AIA filters, viz., 94, 131, 335, 211, and 171, and compared them with the observed light curves in Figs.~\ref{fig:aiarlc00}, \ref{fig:aiarlc07}, \& \ref{fig:aiarlc26}.

The average model intensities of the events in each AIA filter, corresponding to input parameters selected using both methods are reported in Table~\ref{table:aiaintense}, and are compared with the measured intensities from \cite{srividya}. We also show the range in the calculated intensities that arises due to possible systematic uncertainty in the event duration. This is done by computing average intensities when the duration of heating is $\frac{1}{2}\times$ and 2$\times$ the most suitable value.  These results demonstrate that the observed intensities in AIA~193~{\AA} filter are robustly modeled, and where they differ for other filters, point to limitations in the plasma temperature reconstructions with {\ebtel}.

We note that {\ebtel} also allows the inclusion of non-thermal particle flux within the simulation. However, we find that even requiring as much as half of the total energy budget be dissipated by non-thermal particles (for loop parameters relevant to our study, as in \S\ref{sec:analysis}), the light curves change by $<$5\%. Additionally the presence of non-thermal particles worsens the agreement between simulated and observed EM weighted temperatures. Therefore, we have switched off this option in the code.

In order to study the energetics of brightenings of this class, we now look into the {\ebtel} simulations to understand how the energy transfer processes operate.  We show in Figs.~\ref{fig:energetics00}, \ref{fig:energetics07}, \& \ref{fig:energetics26}, the conduction loss (blue), radiation loss (green), enthalpy (red), and the heating rate (black) for each brightening simulated using input parameters obtained from method 1 (panels { [a-b]}) and method 2 (panels { [c-d]}). We note that for the sake of visibility, loss curves are shown only for the most suitable $t_{dur}$. In order to demonstrate both the absolute and relative magnitudes of the energy losses, the contribution of background heating is included in the upper panels of these figures, and are excluded in the lower panels. Note that the values are as computed for the full loop, not just the semi-loop as computed by {\ebtel}.  We next discuss the results for each event in sequence.

\begin{table}
\centering
\caption{Input parameters}\label{table:inputparam}
\begin{tabular}{|c| c| c| c| c| c| c|}
\hline  
\hline 

Index	& \multicolumn{2}{c|}{Energy (E [log(ergs)])} & \multicolumn{2}{c|}{Half Length (L [10$^{8}$ $cm$])} & \multicolumn{2}{c|}{Duration of heating (t$_{dur}$ [$s$])}\\ 

\cline{2-7}
  & Method 1 &  Method 2 &Method 1 & Method 2 & Method 1 & Method 2 \\
  & \tiny{(bkg. modelled)} &  \tiny{(bkg. not modelled)} & \tiny{(bkg. modelled)} & \tiny{(bkg. not modelled)} & \tiny{(bkg. modelled)} & \tiny{(bkg. not modelled)} \\  
\hline  
00 & 23.0 &  23.2  & 1.00  & 0.65 & 60 & 40 \\

07 & 23.1 &  23.3  & 0.90  & 0.65 & 70 & 60 \\

26 & 22.8 &  23.1  & 1.1  & 0.65 &  60 & 70 \\

\hline     

\end{tabular}
\end{table}
\subsection{Modelling the Dynamics of BR-00}\label{subsec:br00}
\subsubsection{Method 1: Background modelled}\label{br00m1}

We find that a loop of half length 1~Mm, with 10$^{23}$~ergs deposited within 60~s mimics the observed rise times, peak intensity, and approximate duration. The pre-event ambient background heating, designed to match the base intensity level in AIA~193, maintains plasma at temperature of 0.92~MK and density 7.8$\times 10^{9}$~cm$^{-3}$, which is similar to the cooler loops found in active regions \citep{Ghosh_2017}. The {\ebtel} simulation is run for 60~s before a triangular heat pulse is applied for a duration $t_{dur}$=60~s. We follow the evolution of the plasma during and after this pulse and use the DEMs obtained from the simulation to predict the intensity light curves for all AIA channels.

\begin{figure*}[h!]
\centering
\includegraphics[width=0.49\textwidth]{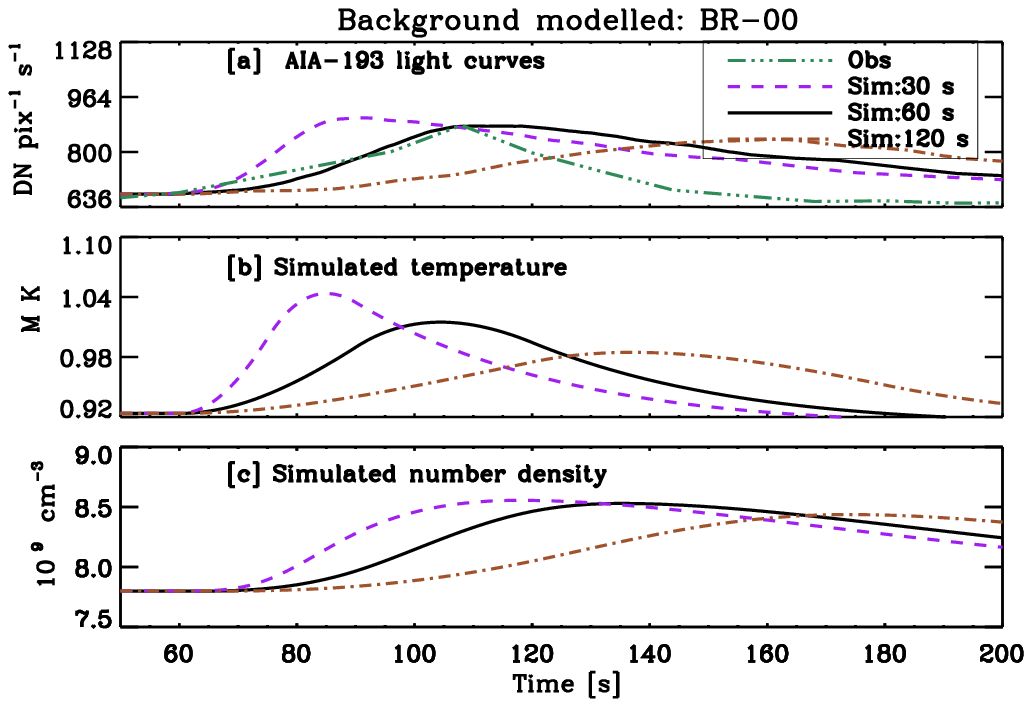} 
\includegraphics[width=0.49\textwidth]{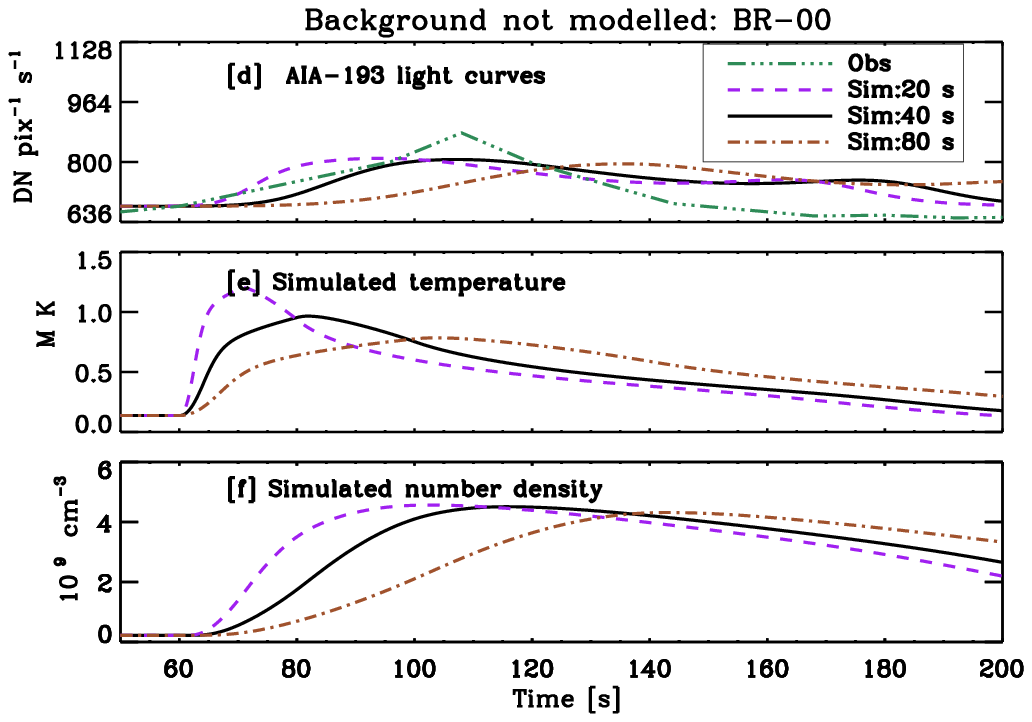} 
\caption{Data and {\ebtel} simulations of brightening BR00 using input parameters set using 
method 1, which uses transient + background for determining input parameters (left column) and method 2 which uses only transient for determining input parameters (right column). Panels a \& d show observed and simulated light curves as labelled. Panels (b, e) and (c, f) show the simulated plasma temperature and density respectively.
The solid curves represent model curves obtained using the parameters in Table~\ref{table:inputparam}, the pink dashed and brown dot-dashed curves represent model curves made for heating durations $\frac{1}{2}\times$ and $2\times$ the nominal, and the green triple-dot-dashed curves represent observed light curves.
} \label{fig:br00res}

\end{figure*}
\begin{figure}[h!]
\centering
\includegraphics[width=0.49\textwidth]{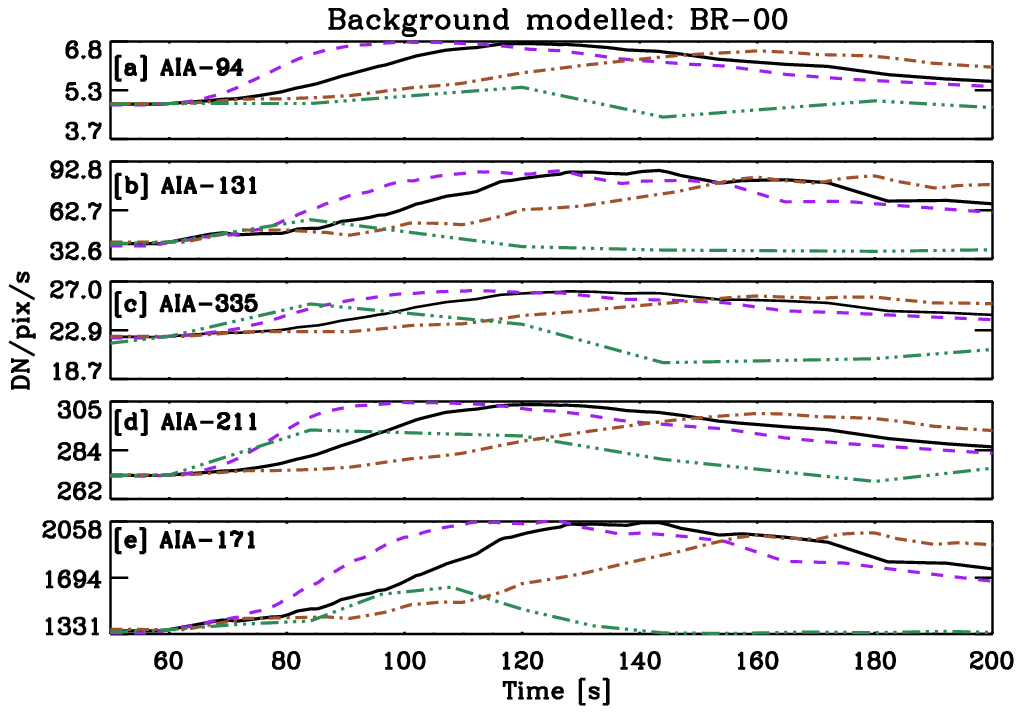} 
\includegraphics[width=0.49\textwidth]{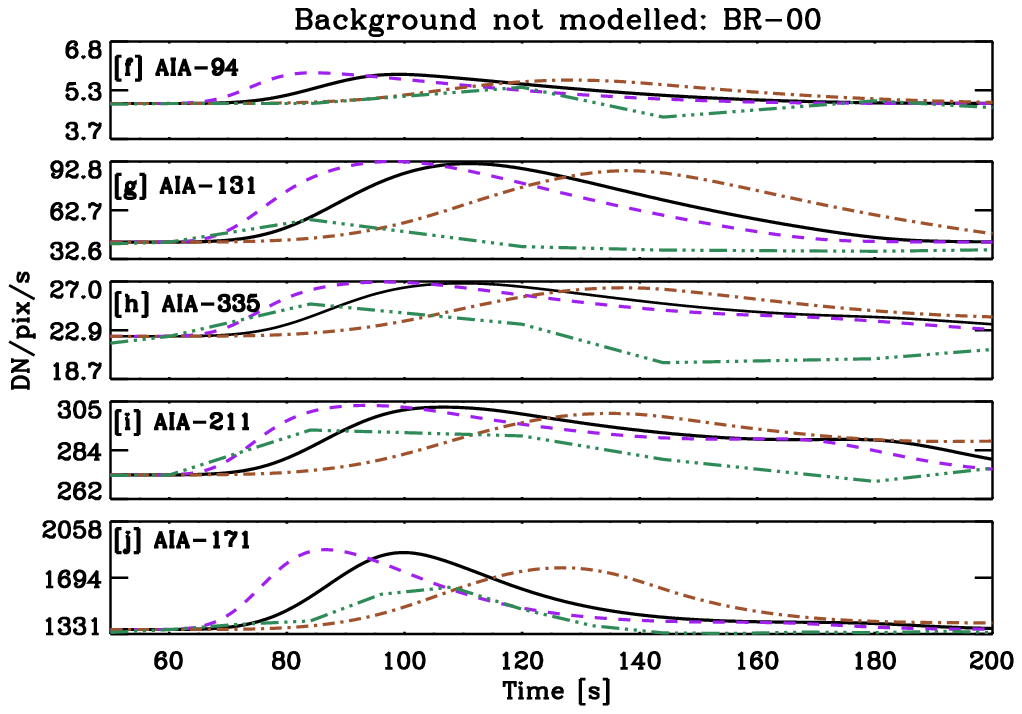} 
\caption{Observed and simulated light curves using input parameters set using 
method 1 (which models the background and the transient together; left column) and method 2 (which considers the background separately from the transient; right column) for BR-00 for AIA channels 94 (panel a[f]), 131 (panel b[g]), 335 (panel c[h]), 211 (panel d[i]) and 171 {\AA} (panel e[j]) as labelled. The colors and line styles 
correspond to those in Fig.~\ref{fig:br00res}. The simulated light curves obtained using method 1, where input parameters have been selected using transient + background  (left column) and method 2, where input parameters have been selected using only transient (right column) have been increased or decreased by a constant offset in each filter to match the observed background level. Note that the best durations chosen for the two methods are different, and therefore the 1/2x and 2x durations also differ. }\label{fig:aiarlc00}

\end{figure}
\begin{figure*}[h!]
\centering
\includegraphics[width=0.49\textwidth]{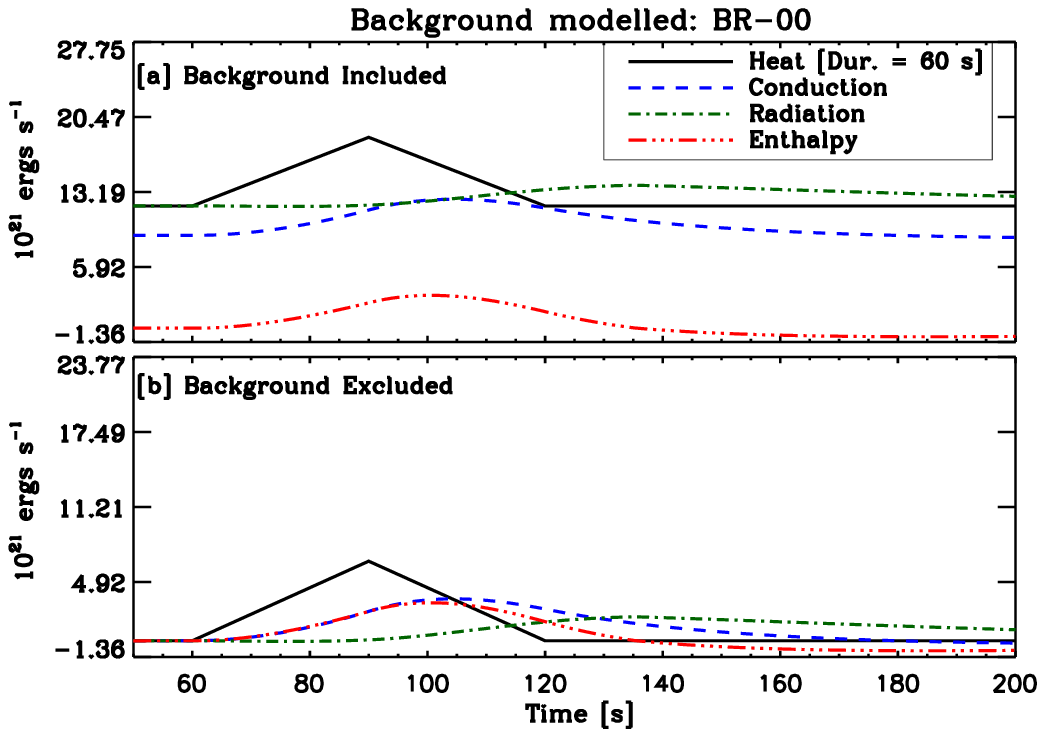} 
\includegraphics[width=0.49\textwidth]{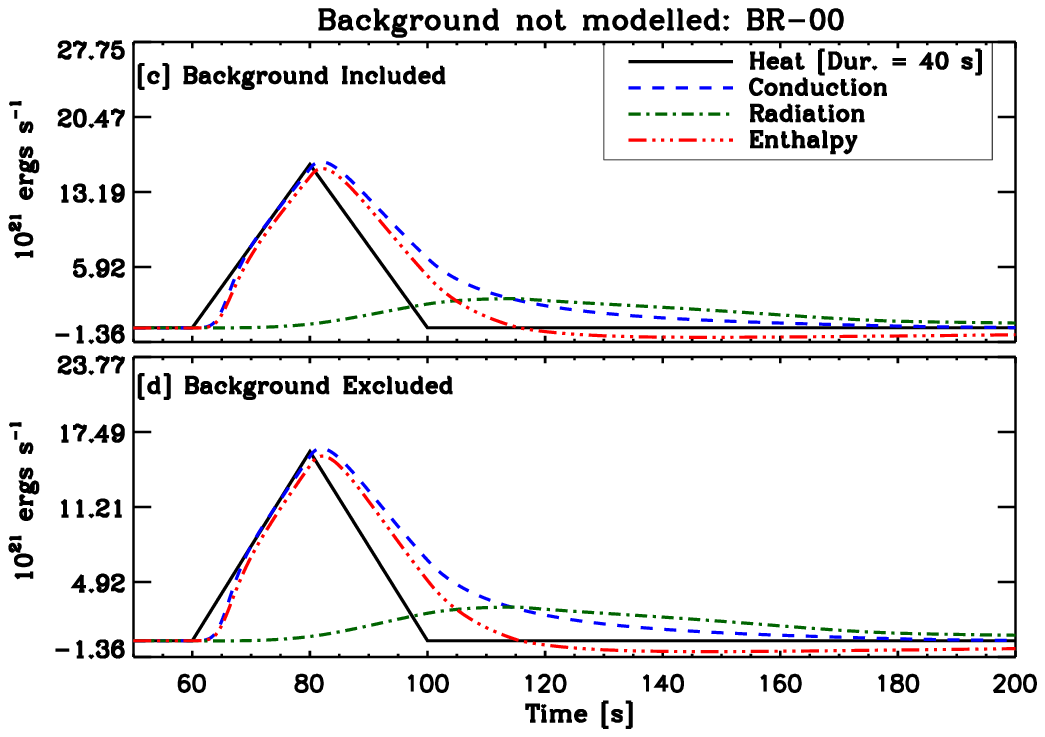} 
\caption{Energy loss and transfer components from {\ebtel} simulations designed to match BR-00. 
The panels in the left column correspond to method 1 (background included in the modeling), and those in the right column to method 2 (backgound treated as separate from the transient). The heat input (solid curves), conduction (blue dashed curves), radiation loss (green dot-dashed curves), and enthalpy (red triple-dot-dashed curves) are shown.
The top panels show curves that include the background while bottom panels are shown with background values subtracted.
Note that the radiation loss term includes contributions from both the corona and the transition region.} \label{fig:energetics00}
\end{figure*}

Simulation light curves of AIA~193 intensity, plasma temperature, and plasma density are shown in the left column of Fig.~\ref{fig:br00res}, for the nominal heating duration of $t_{dur}$=60~s (black curve), as well as for 30~s (red dashed curves) and 120~s (red dot-dashed curves). As expected, the simulated intensity light curves match the achieved intensity and duration of the observed light curve. Note that the peak intensity, temperature, and density are all correlated with the heating duration; this is due to more impulsive events heating the plasma on a smaller timescales, leading to a sharper rise in temperature and consequently a higher density due to stronger evaporation.

We also show the simulated intensity light curves for the other AIA filters in the left column of Fig.~\ref{fig:aiarlc00}.
In these plots the simulated light curves have been 
offset 
by a case specific 
value 
to match the observed pre-event background intensity level. 
The background level has been offset
from that predicted based on matching to AIA\,193
(see Table~\ref{table:offsets}) to match the model light curve intensities to the pre-event observed intensities
for each filter.
{As seen above for the 193\,\AA\ filter, the anti-correlation between peak intensity and heating duration persists for all filters.} 
It is seen that 
observed intensities 
in all AIA filters except 94\,\AA\ peak before the synthetic light curves
for the most suitable heating duration $t_{dur}=60$~s. The agreement is best for 211\,\AA\ filter and worst for AIA\,171 and 131\,\AA\ filters.

Next, we compute the average intensities of the simulated light curves in each of the AIA filters.  The light curves are integrated over a lifetime as determined from when the AIA~193 \AA~light curve drops to 5\% of the peak intensity, and divided by the lifetime to obtain a comparison with the observed average intensities (see columns 2 and 3 of Table~\ref{table:aiaintense}). The simulated intensities match observations well for AIA~193, though differ by factors ranging from $0.8\times$ (AIA~335) to $3.1\times$ (AIA~171) in other filters.  This is a remarkable correspondence with observations considering the limitations of the 0-D monolithic loop system we consider.  

The energy loss and transfer terms (conduction and radiation losses, enthalpy, and heat input) for the simulation with the nominal heating duration is shown in Fig.~\ref{fig:energetics00}. The upper panel includes the contribution from the ambient background, and the lower panel isolates the effects due only to the brightening event. Note that enthalpy (red triple-dot-dashed curves; Eq.~\ref{eq:enthalpy}) can be positive or negative depending whether plasma is flowing into or out of the corona. As expected, conduction losses (blue dashed curve; Eq.~\ref{eq:condloss}) dominate at the beginning (until $\approx$70~s) of the event, and radiation losses (green dot-dashed curve; Eq.~\ref{eq:radloss}) dominate at later times. The enthalpy into the corona keeps pace with conduction loss in the early phase (for $\approx$40~s), and drops off $\approx$10~s before conduction does. The enthalpy reverses sign at approximately the same time that radiation becomes the dominant loss mechanism. The total time integrated conduction loss from the coronal loop (which is eventually radiated) is $3.0\times10^{23}$ ergs. The net enthalpy is positive (into corona) and is equal to $4.5\times10^{22}$ ergs. This is an order of magnitude less than the heating function. Notice that the radiation loss drops below the ambient background level at the beginning of the event, as does conduction loss about 120~s after the heat pulse. These represent small perturbations in the ambient coronal structure, and do not have any effect on the energetics.

\subsubsection{Method 2: Background not modelled}\label{br00m2}

An impulse with a triangular profile dissipating a total of $10^{23.2}$~ergs in 40~s is best suited for a loop of half length of 0.65~Mm subjected to uniform background heating of $10^{-4}$~ergs~$cm^{-3}$~s$^{-1}$ (see~\ref{subsubsec:analysism2p1}). The plots in the right column of Fig.~\ref{fig:br00res} display the observed and simulated light curves for 193\,{\AA} (panel d), plasma temperature (panel e), and density (panel f). Different colors belong to $t_{dur}$ (black), $\frac{1}{2}\times t_{dur}$ (purple) and $2\times t_{dur}$ (brown) as labelled. The simulated background intensity is almost three orders of magnitude smaller than the peak simulated values and hence negligible. Note that to bring the background level of the synthetic light curves to that of observed, we have added a case specific offset (see~\ref{table:offsets}). As expected, the peak values of temperature, density, and intensities are correlated with the duration of heating. Even though the initial density and temperature in this case are an order of magnitude lower than the values obtained by simulations using method 1, the peak values of temperature match in both cases and peak density produced by method 2 is less than that produced by method 1 by a factor $\lesssim 2$. 

We have also obtained the synthetic light curves for the other channels of AIA and plotted them after 
adding case specific background levels,
in the right column of Fig.~\ref{fig:aiarlc00}. The colors have the same meaning as those in the right column of Fig.~\ref{fig:br00res}. For comparison, we have also over-plotted the observed light curves for the corresponding AIA channels.

We plot the various energy loss and transfer terms, both including and excluding the background contribution in the right column of Fig.~\ref{fig:energetics00} (see panels c \& d). The qualitative feature of an initial conduction-dominated loss is the same as that obtained using method 1 and is shown in panels a \&~b of Fig.~\ref{fig:energetics00}. Moreover, the time at which enthalpy changes its sign is also 
approximately 
the same
as
where radiation loss starts dominating conduction loss, similar to method 1. However, we note that enthalpy plays a more important role in method 2 than in method 1. The time integrated intensities obtained from synthetic light curves in all filters show a better correspondence with those obtained from the observed light curve, in particular for 171\,{\AA}, which shows the highest discrepancy.

\begin{figure*}[h!]
\centering
 \includegraphics[width=0.49\textwidth]{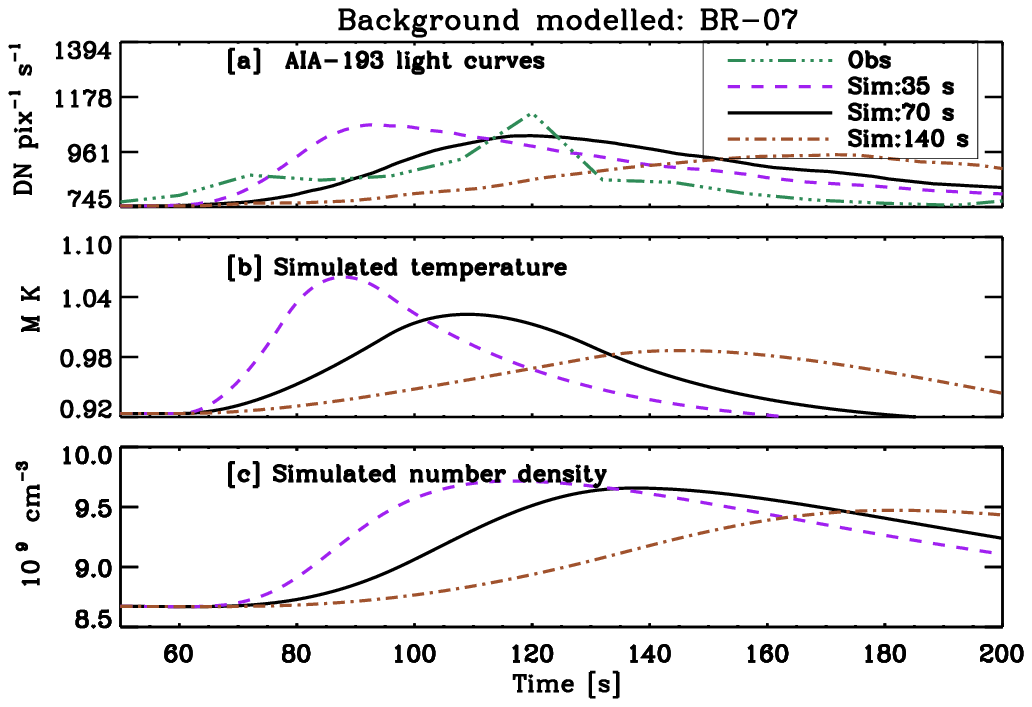}
 \includegraphics[width=0.49\textwidth]{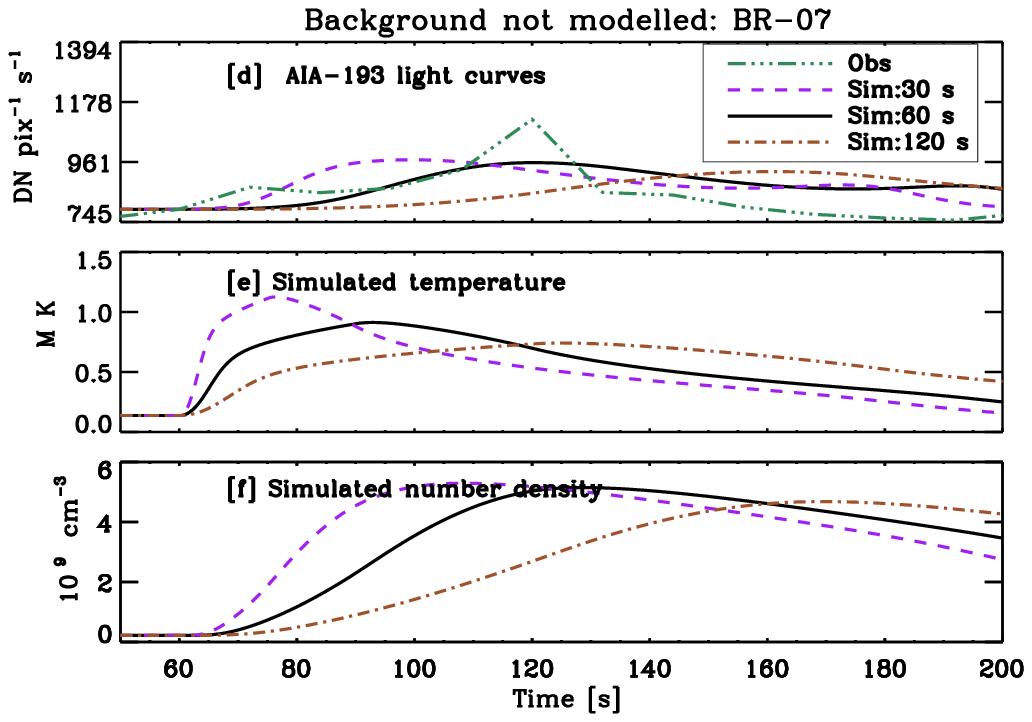}
   \caption{Same as in Fig.~\ref{fig:br00res} but for BR-07.}\label{fig:br07res}
\end{figure*}
\begin{figure*}[h!]
\centering
 \includegraphics[width=0.49\textwidth]{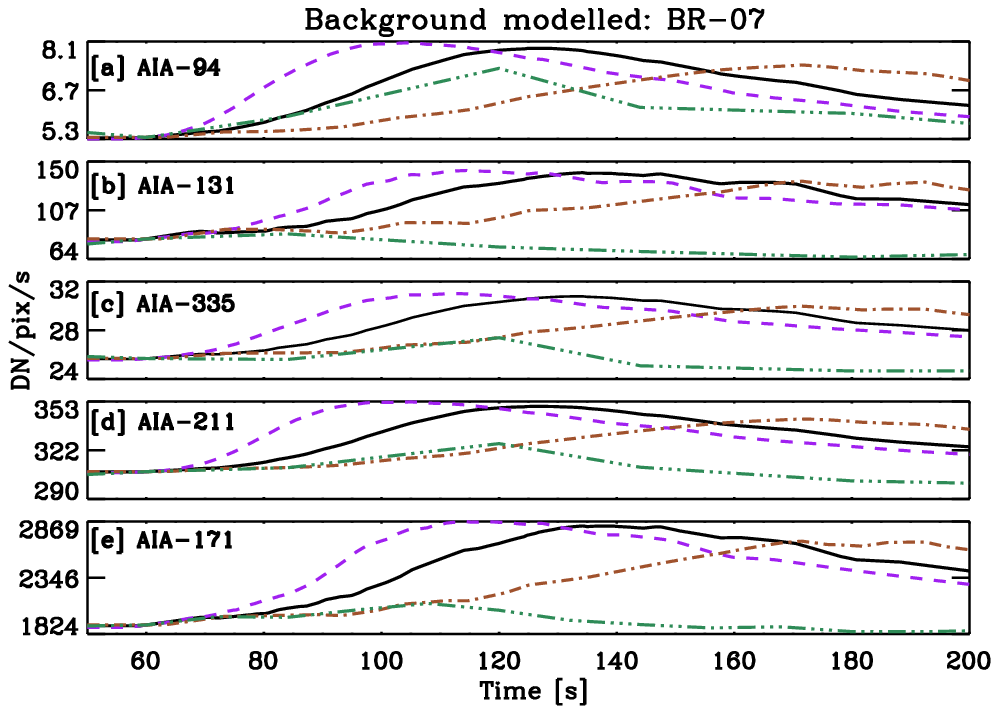}
 \includegraphics[width=0.49\textwidth]{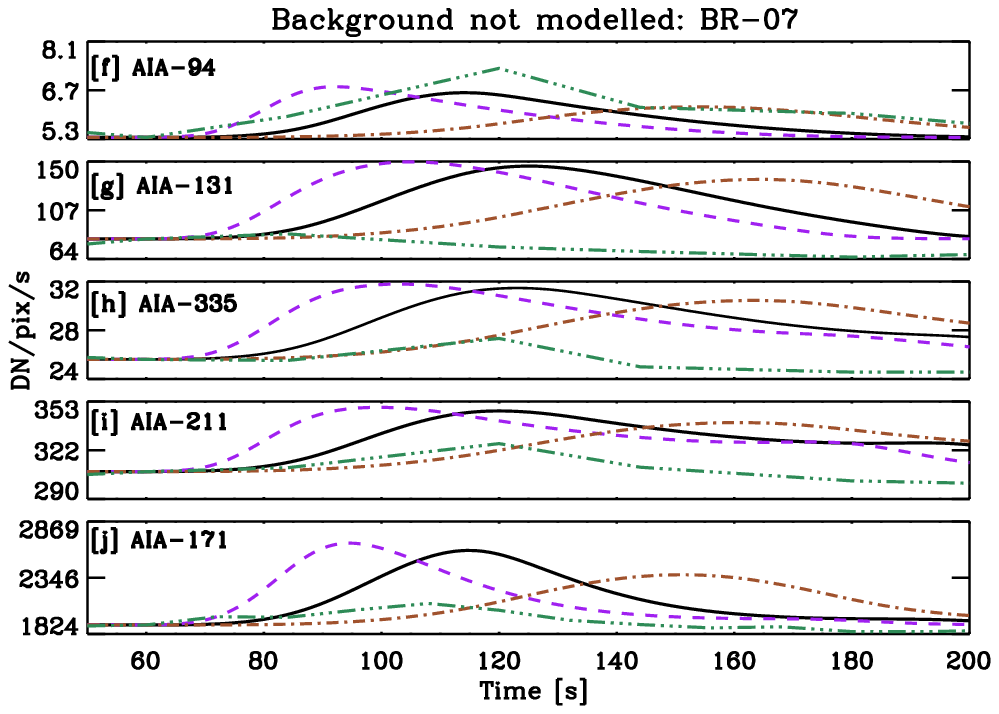}
 \caption{Same as Fig.~\ref{fig:aiarlc00} but for BR-07.  }\label{fig:aiarlc07} 
 
\end{figure*}
\subsection{Modelling the Dynamics of BR-07}\label{subsec:br07}
\subsubsection{Method 1: Background modelled}\label{br07m1}

 \begin{figure*}[h!]
 \centering
 \includegraphics[width=0.45\textwidth]{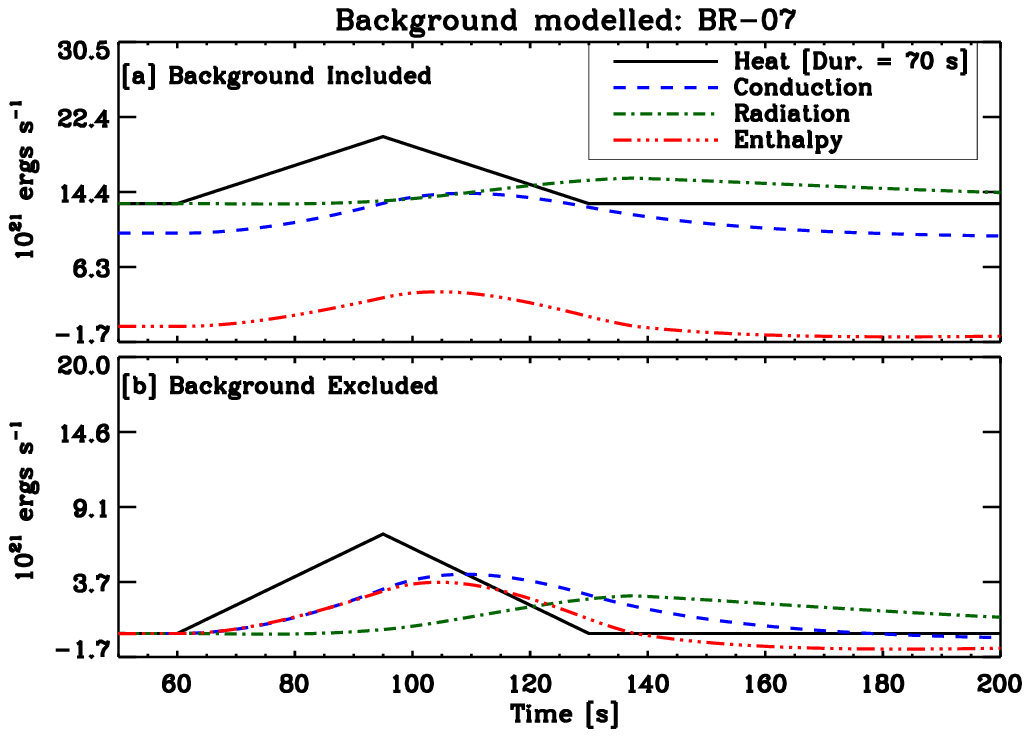} 
  \includegraphics[width=0.45\textwidth]{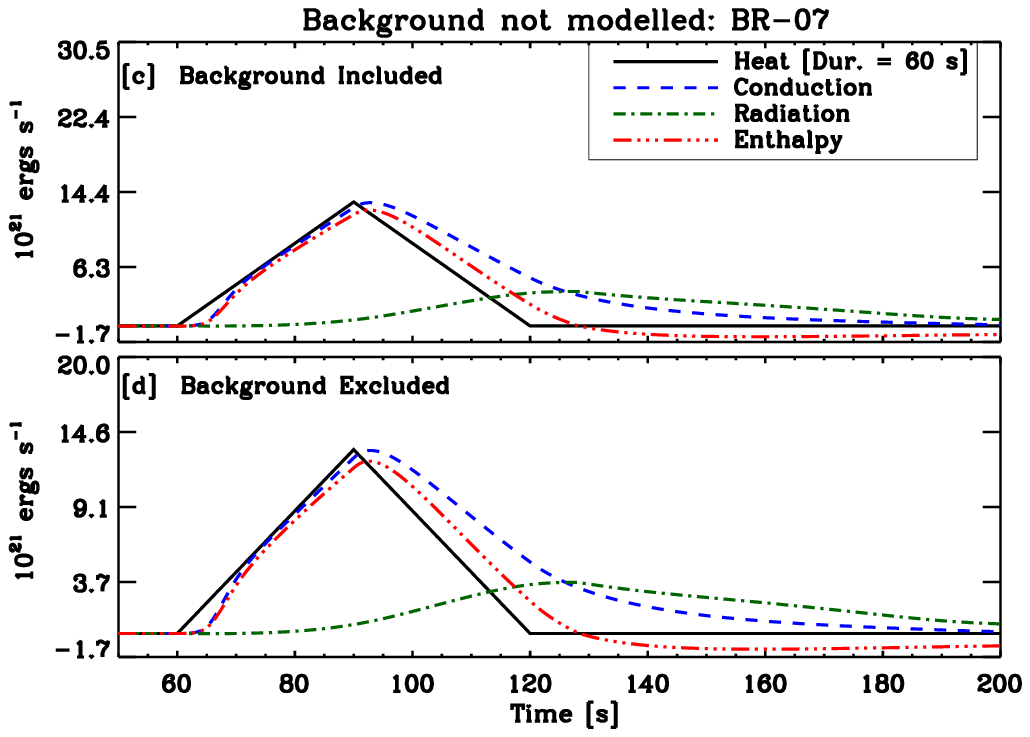}
 \caption{Same as in Fig.~\ref{fig:energetics00} but for BR-07.}\label{fig:energetics07}
\end{figure*}

We follow the same procedure to analyze BR-07 here as we did for BR-00 in \S\ref{br00m1}.
A loop of half-length of 0.9 Mm, with $10^{23.1}$ ergs deposited in $t_{dur} = $ 70 s mimics the observed rise times, peak intensity and approximate duration. The background heating used for matching the base intensity levels in AIA 193 maintains plasma at a temperature and density of 0.92 MK and $8.7\times 10^{9}$ cm$^{-3}$. The simulation is run for 60 s before a triangular heat pulse is applied for a duration of $t_{dur}$=70 s. The evolution of plasma obtained from simulations were used to predict light curves in all channels. Simulation light curves of AIA 193 intensity, temperature and density of plasma for $t_{dur}$, $2\times t_{dur}$ and $\frac{1}{2}\times t_{dur}$ along with observed AIA 193 light curve are shown in Fig.~\ref{fig:br07res}. Simulated light curves in remaining filters (after increasing or decreasing by case specific offsets) are shown in Fig.~\ref{fig:aiarlc07}. The energy loss and transfer terms for simulations with nominal heating duration i.e. 70 s in this case, is shown in Fig.~\ref{fig:energetics07}. The scheme used in the figures are same as that used for BR-00. 

The peak intensity in all filters, temperature and density are largest for most impulsive heating. The simulated intensities match observations well for AIA 193. It differs in other filters by factors between 1.4$\times$ (AIA 94) and 5.5$\times$ (AIA 171).  The event has an initial conduction dominated cooling phase lasting for $\approx$ 70 seconds followed by radiation dominated cooling phase at later times. Enthalpy starts dropping $\approx$ 10 s before conduction. The total time integrated conduction loss from the coronal loop is $3.7\times10^{23}$ ergs. The net enthalpy is positive (into corona) and is equal to $5.1\times10^{22}$ ergs.

\subsubsection{Method 2: Background not modelled}\label{br07m2}

We follow the same procedure to analyze BR-07 as in \S\ref{br00m2}.
A heating event having a triangular profile, which dissipates $10^{23.3}$ $ergs$ in 60 $s$ is best suited for a loop of half length of 0.65 Mm subjected to uniform background heating of $10^{-4}$ $ergs~cm^{-3}~s^{-1}$.  A case specific offset has been added to all the simulated light curves to make the observed and simulated pre-event intensities equal (see Fig.~\ref{fig:br07res} and \ref{fig:aiarlc07}). All features including an initial conduction dominated cooling of corona, enthalpy changing sign 
approximately
when radiation starts dominating, are qualitatively the same as that of BR-00 (see Fig.~\ref{fig:energetics07}). The average intensity in AIA\,171 filter agrees better with the observed value.

\begin{figure*}[h!]
\centering
 \includegraphics[width=0.49\textwidth]{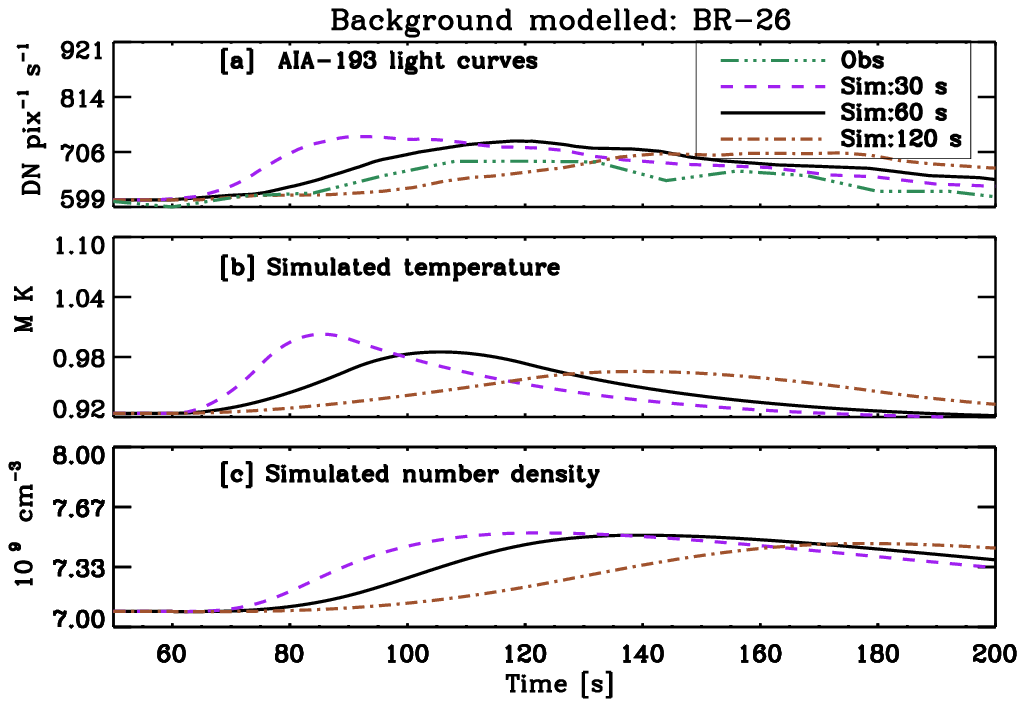} 
  \includegraphics[width=0.49\textwidth]{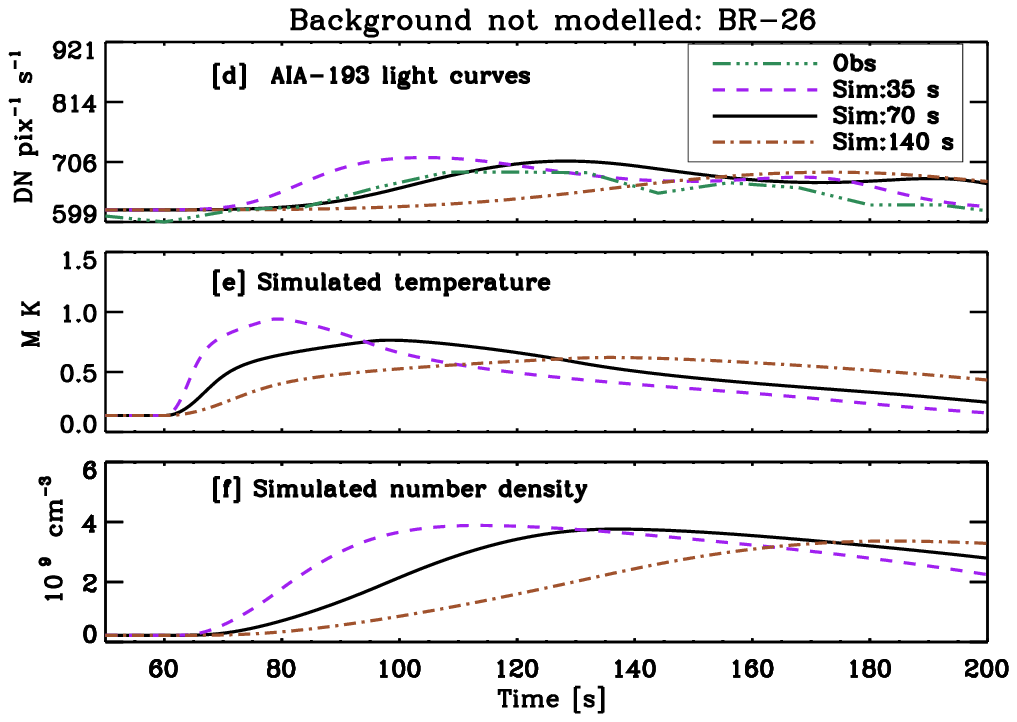} 
   \caption{Same as Fig.~\ref{fig:br00res} but for BR-26.}\label{fig:br26res}
\end{figure*}
\begin{figure*}[h!]
\centering
 \includegraphics[width=0.49\textwidth]{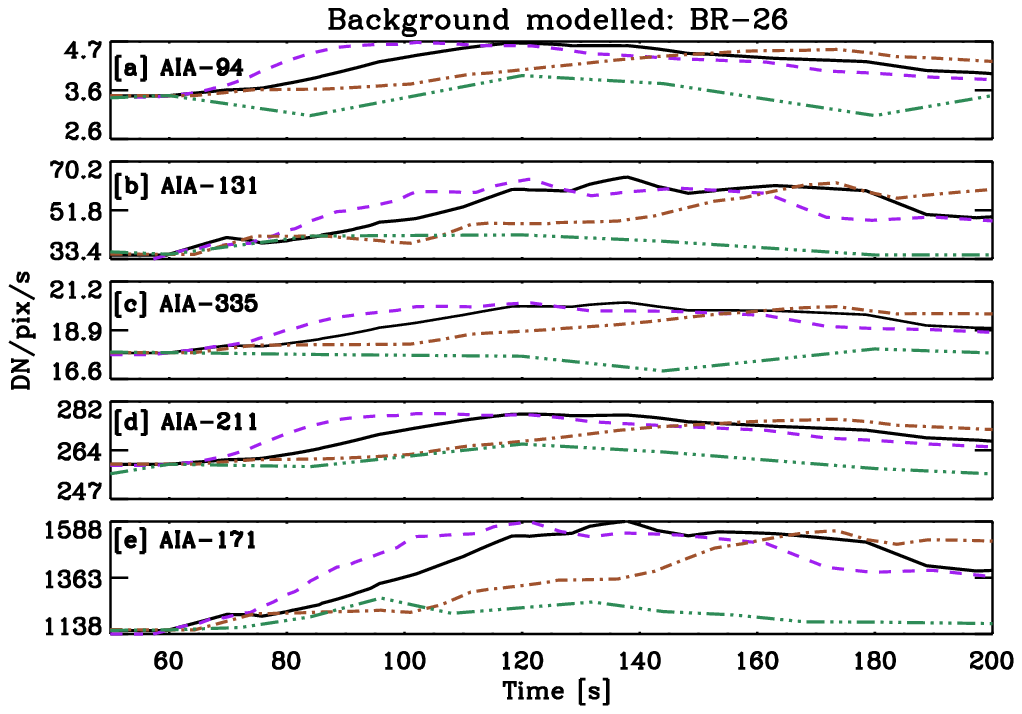} 
  \includegraphics[width=0.49\textwidth]{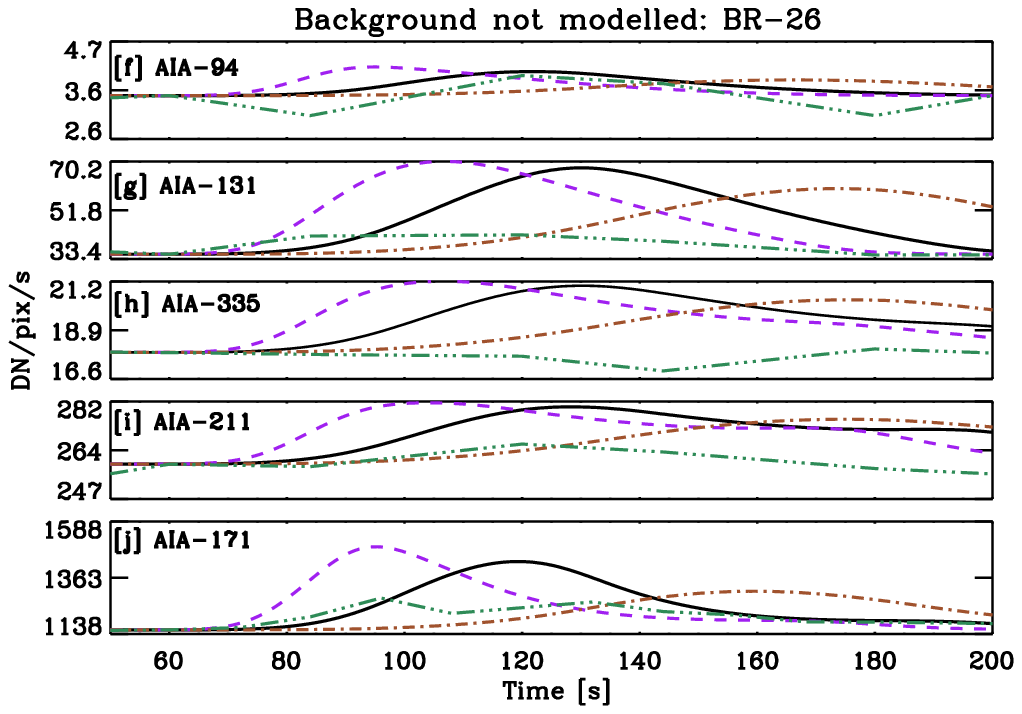} 
 \caption{Same as Fig.~\ref{fig:aiarlc00} but for BR-26.}\label{fig:aiarlc26}

\end{figure*}

 \begin{figure*}[h!]
 \centering
 \includegraphics[width=0.49\textwidth]{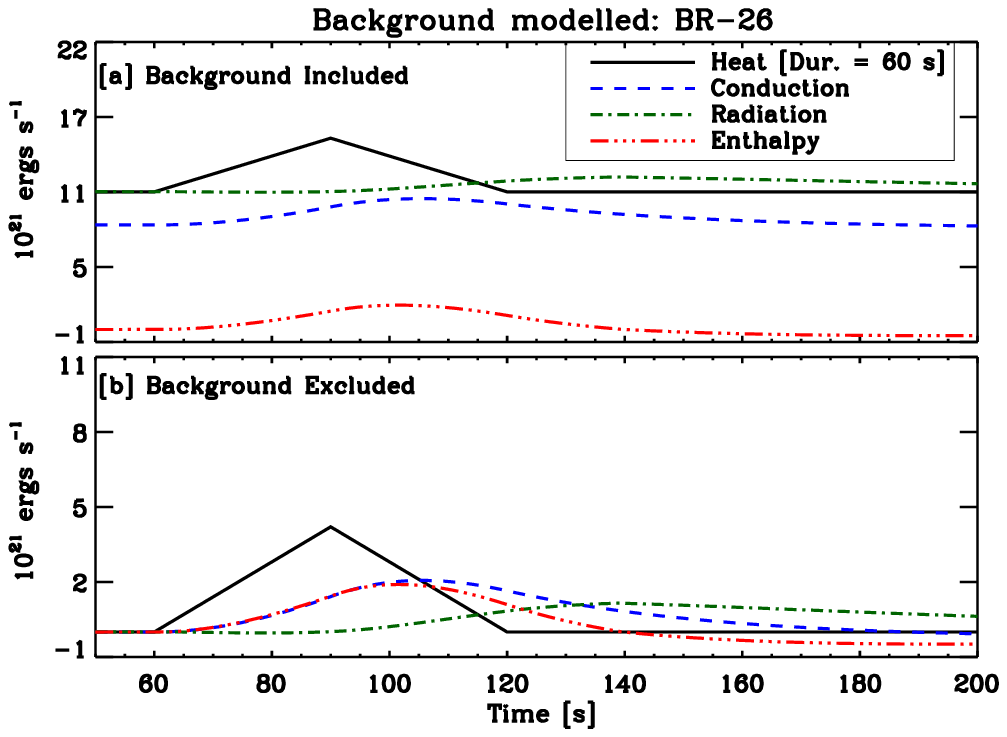} 
  \includegraphics[width=0.49\textwidth]{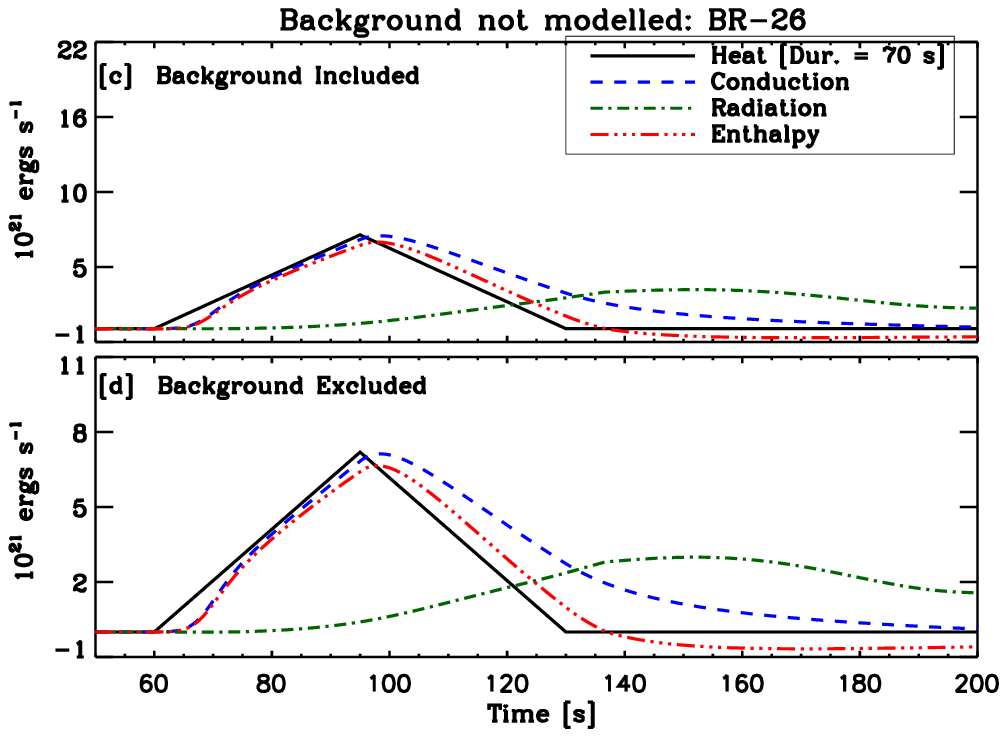} 
 \caption{Same as Fig.~\ref{fig:energetics00} but for BR-26.}
 \label{fig:energetics26}
\end{figure*}
\subsection{Modelling the Dynamics of BR-26}\label{subsec:br26}
\subsubsection{Method 1: Background modelled}
We follow the same procedure to analyze BR-26 here as for BR-00 in \S\ref{br00m1}.
A loop of half-length of 1.1 Mm, with $10^{22.8}$ ergs deposited in $t_{dur} = $ 60 s mimics the observed light curves. The background heating used for matching the base intensity levels in AIA 193~\AA~maintains plasma at a temperature and density of 0.92 MK and $7.1\times 10^{9}$ cm$^{-3}$. Simulated light curves of AIA 193 intensity, temperature and density of plasma for $t_{dur}$, $2\times t_{dur}$ and $\frac{1}{2}\times t_{dur}$ along with observed AIA 193 light curve are shown in Fig.~\ref{fig:br26res}. Simulated light curves in remaining filters (after increasing or decreasing by case specific offsets) are shown in Fig.~\ref{fig:aiarlc26}.   The average intensities in filters apart from AIA-193 differ by factors ranging between 1.2$\times$ (AIA 94) and 3.6$\times$ (AIA 171). The energy loss and transfer terms for simulations with nominal heating duration i.e. 60 s in this case, is shown in Fig.~\ref{fig:energetics26}.  The scheme used in the plots are same as that used for BR-00. Energy loss and transfer evolution trends are similar to those for BR-00 and BR-07. The total time integrated conduction loss from the coronal loop is $1.9\times10^{23}$ ergs. The net enthalpy is positive (into corona) and is equal to $3.5\times10^{22}$ ergs.

\subsubsection{Method 2: Background not modelled}
We follow the same procedure to analyze BR-26 as detailed in \S~\ref{br00m2}.
A heating event having a triangular profile, which dissipates $10^{23.1}$~ergs in 70 $s$ is best suited for a loop of half length of 0.65 Mm subjected to uniform background heating of $10^{-4}$~ergs~cm$^{-3}$~s$^{-1}$. 
A case specific offset has been added to all the simulated light curves to make observed and simulated background intensities equal (see Figs.~\ref{fig:br26res} and \ref{fig:aiarlc26}). Once again we see an initial conduction dominated cooling of corona, enthalpy changing sign 
approximately 
when radiation starts dominating, are qualitatively the same as that of BR-00 (see Fig.~\ref{fig:energetics26}). The average intensity in AIA-171 filter agrees better with the observed value.

\begin{table} 
\centering
\caption{Observed and simulated average Intensities in DN~pix$^{-1}$~s$^{-1}$ over AIA-193 lifetime of brightenings. The values in the parenthesis cover the range of average intensities if the duration in which energy is deposited in  half and double the time duration selected as the most suitable input parameter.} \label{table:aiaintense}

\begin{tabular}{|c| c| c| c| c| c| c| c| c| c|}
\hline  
\hline 
AIA	& \multicolumn{3}{c|}{BR-00 (DN~pix$^{-1}$~s$^{-1}$)} & \multicolumn{3}{c|}{BR-07 (DN~pix$^{-1}$~s$^{-1}$)} & \multicolumn{3}{c|}{BR-26 (DN~pix$^{-1}$~s$^{-1}$)}\\ 
\cline{2-10}
Filter & Obs. &  \multicolumn{2}{c|}{Sim.} & Obs. & \multicolumn{2}{c|}{Sim.} &  Obs. & \multicolumn{2}{c|}{Sim.}  \\
\cline{3-4}\cline{6-7}\cline{9-10}
  &   &  Method 1 & Method 2 &   & Method 1 & Method 2 &    & Method 1 & Method 2  \\
  &   &  \tiny{(bkg. modelled)} & \tiny{(bkg. not modelled)} &   & \tiny{(bkg. modelled)} & \tiny{(bkg. not modelled)} &    & \tiny{(bkg. modelled)} & \tiny{(bkg. not modelled)}  \\
  
\hline
AIA-094 & 0.50 & 0.92 & 0.31 & 0.89 & 1.2 & 0.43 & 0.40 & 0.48 & 0.17 \\

  &   & [0.78-1.0]  &  [0.25-0.34]   &  & [0.96-1.3]  & [0.31-0.49] &  & [0.41-0.61] &    [0.12-0.21] \\
\hline  
AIA-131 & 7.3 & 24.0 & 19.7 & 9.1 & 31.7 & 25.9 & 4.6 & 13.9 & 12.2 \\

   &   & [20.9-26.2] & [16.9-21.45] &   & [25.6-35.0] & [20.4-29.3]  &   & [12.0-16.9] & [9.3-14.0] \\
\hline  
AIA-171 & 104.3 & 376.5 & 146.9 & 88.6 & 489.6 & 207.2 & 60.1 & 214.0 & 84.2 \\

   &   & [325.9-410.3] & [124.2-155.4] &   & [401.3-537.1] & [153.5-228.3] &   & [183.3-260.6] & [53.8-97.9] \\
\hline     
AIA-193 & 83.8 & 85.6 & 67.3 & 101.4 & 112.2 & 83.0 & 39.7 & 41.5 & 44.0 \\

  &   & [70.1-95.1] & [56.8-73.9]  &   & [87.8-127.7]  & [64.6-95.4] &   &  [35.3-55.6] & [33.1-51.0] \\
\hline    
AIA-211 & 12.0 & 15.2 & 16.4 & 8.7 & 19.9 & 20.0 & 5.2 & 7.8 & 11.0 \\

   &   & [12.7-16.8]  & [14.0-17.9] &   & [15.8-22.4]  & [15.7-22.7]  &  &  [6.7-10.1] & [8.3-12.7] \\
\hline     
AIA-335 & 2.8 & 2.0 & 2.3 & 1.1 & 2.6 & 2.9 & 0.64 &  1.1 & 1.6 \\

   &   & [1.7-2.2] & [2.0-2.5] &   & [2.1-2.9] & [2.3-3.2] &   & [0.94-1.4] & [1.2-1.8] \\
 \hline   
\end{tabular}
\end{table}

\begin{table} 
\centering
\caption{Background levels used to compare observed and model light curves in various filter in Figs.~\ref{fig:br00res},\ref{fig:aiarlc00},\ref{fig:br07res},\ref{fig:aiarlc07},\ref{fig:br26res},\ref{fig:aiarlc26}.  The values listed for Method 1 are {\sl offsets} relative to the simulated value derived by comparison with AIA\,193 light curves (see text).}
\label{table:offsets}

\begin{tabular}{|c| c| c| c| c| c| c|}
\hline  
\hline 
AIA	& \multicolumn{2}{c|}{BR-00 (DN~pix$^{-1}$~s$^{-1}$)} & \multicolumn{2}{c|}{BR-07 (DN~pix$^{-1}$~s$^{-1}$)} & \multicolumn{2}{c|}{BR-26 (DN~pix$^{-1}$~s$^{-1}$)}\\ 
\cline{2-7}
 Filter &     Method 1 & Method 2 &    Method 1 & Method 2 &     Method 1 & Method 2  \\
  &    \tiny{(bkg. modelled)} & \tiny{(bkg. not modelled)} &    \tiny{(bkg. modelled)} & \tiny{(bkg. not modelled)} &     \tiny{(bkg. modelled)} & \tiny{(bkg. not modelled)}  \\
  
\hline
AIA-094   & -2.1   &  4.8  &  -0.95  &  5.3  &   -2.8 &  3.5  \\
\hline
AIA-131   & -176.0   & 43.0   & -116.9   & 81.7   & -163.4   & 35.2   \\
\hline
AIA-171   & -1908.6   & 1359.8   & -1056.2   &  1908.0  &  -1809.8  & 1154.5   \\
\hline 
AIA-193   & 0   & 679.0   &  0  & 790.7    &  0  &  620.6  \\
\hline 
AIA-211   &  +131.4  & 272.6   & +179.5   & 307.7   &  +131.3  &  259.5  \\
\hline 
AIA-335   &  +2.9  & 22.3   & +7.9   &  25.6  &  +0.18  &  17.8  \\
 \hline   
\end{tabular}
\end{table}

\subsection{Comparison of conductive flux to the radiative flux}\label{comp}
\begin{figure}[h!]
\centering
 \includegraphics[width=0.49\textwidth]{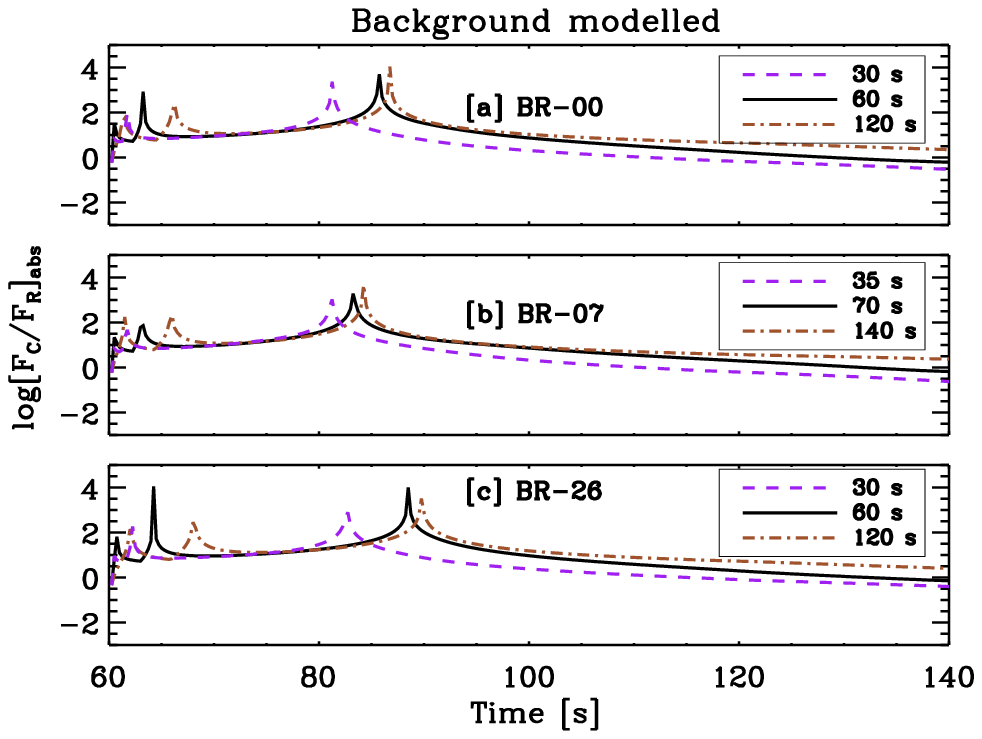}
  \includegraphics[width=0.49\textwidth]{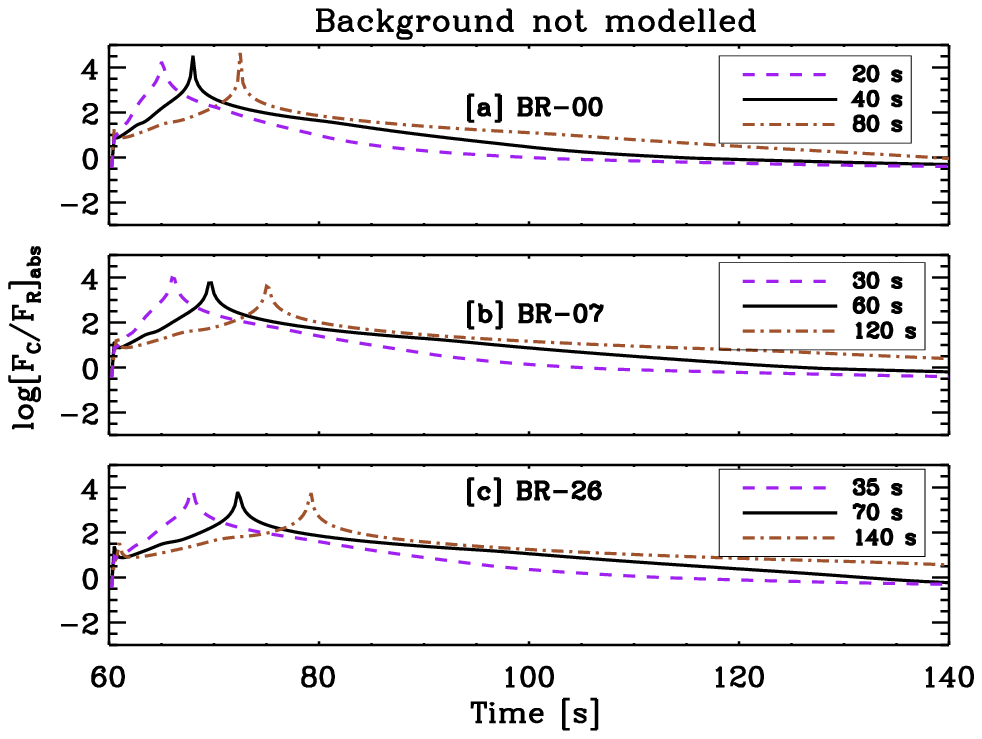}
 \caption{Time evolution of the ratio of absolute values of conduction 
 and radiation loss rates for BR-00 (panel [a]), BR-07 (panel [b]) and BR-26 (panel [c]). The solid black curves are for the optimized duration while the dashed-blue and dashed-dotted red corresponds to the result obtained for half and twice the duration of the heating event as labelled.}
 \label{fig:clifs}
\end{figure}

To quantitatively assess the relative importance of conduction loss over radiation loss during different phases of the brightening, in Fig.~\ref{fig:clifs} we plot the logarithm of the ratio of absolute values of background subtracted conduction and radiation loss rates obtained from method 1 (left column) and method 2 (right column);
the upper panels correspond to BR-00, the middle panels 
to BR-07, and the bottom panels 
to 
BR-26.
The solid black curves 
correspond to the nominal heating 
durations, while the dashed-blue and dashed-dotted red curves correspond to heating durations set to $\frac{1}{2}\times$ and $2\times$ the nominal duration. 
The cusps in the curves (e.g., those present between 80 and 100~s in the left panels and between 60 and 80~s in the right panels) are artefacts of correcting for the background, arising from the radiative loss dropping below and rising above the ambient value (see Fig.~\ref{fig:energetics00}, \ref{fig:energetics07}, \ref{fig:energetics26} and discussion in Section~\ref{subsec:br00}). 
The occurrence of these spikes at different locations for the same brightenings in the
two methods is due to using different input parameters in each case. 
For all the three events, irrespective of how the input parameters are selected, we find an initial phase where conduction is the dominant cooling mechanism in the corona. This phase ends at a simulation time step of $\approx$140~s i.e., 80~s after the heating starts.

\section{ Discussion and Summary }\label{sec:cad}

Hi-C has provided observations of the solar corona in 193~{\AA} at the highest spatial resolution so far. It has presented us with a number of intriguing observations, including those of the faintest  transient brightenings observed \citep[][]{regnier, srividya}. These faint brightenings when studied by multi-wavelength images of AIA, were deemed to be dominated by conduction loss. Here, we have explored coronal energetics using a simplified 0-D description (\ebtel) to simulate the evolution of coronal plasma in loops subject to heat inputs.  

We identify loop sizes and heat inputs that mimic the intensities and durations of the observed brightenings using two methods. In first method we use background intensities for constraining the input parameters and in second method we don't use this information. We use the simulations to study the transfer of energy into and out of the corona.  We focus on three of the simplest brightenings identified by \citep[][]{srividya}, with single unambiguous intensity peaks.  We adopt triangular heating profiles of duration 30-140~s, and find that loop half lengths of $\approx$10$^{8}$~cm and energy deposition of $\approx$10$^{23}$~ergs can generate dynamical intensity profiles that mimic the observed brightenings.

We find that the average brightness of the simulated loops in the AIA~193~{\AA} filter matches the observations well, and are within a factor of 2{--}5.5 in the other AIA filters if input parameters are selected by method 1 (i.e. both transient and background are modelled) and within a factor of 2{--}3 if input input parameters are selected by method 2 (background is not modelled). 
The obtainment of better tally by latter method, can be suggestive of the transient events and background being dynamically distinct. 
In either method the largest discrepancies between the simulated and observed intensities, irrespective of how input parameters have been selected, arises in the AIA~171\AA\ and AIA~131\AA\ filters.

For all three events studied here, we find that conduction is the dominant cooling mechanism in corona in the early phase of the transients. About 80~s after the heat pulse, radiation looses begin to dominate.  We observe that conduction dominated cooling in corona during the early phase of the evolution has also been reported for flares \citep[see e.g.,][]{cargill95}, microflares \citep[see e.g.,][]{gupta}, and are also expected for nanoflares \citep[see e.g.,][]{cargill1994}. Our results show that transient events such as the ones observed in Hi-C are similar in character to microflares and nanoflares, and are likely produced through the same underlying physical processes.  We also note that our simulated plasma temperatures are lower than the observed values. And since conduction loss increases as $T^{3.5}$ while radiative loss decreases in this temperature regime, our assessment of the relative magnitude of conduction loss in the energetics of coronal plasma is an {\sl underestimate}. 

If the background temperature were raised in the model, it necessarily requires an increase in the loop length in order to maintain consistency with observed intensities (see Equation~\ref{eq:background}), resulting in loops of length $\gtrsim$8~Mm and life times $\gtrsim$6$\times$ the observed values.  If the loop lengths were fixed at $\approx$1~Mm, the heating rate increases and causes the predicted model intensities to increase non-linearly, worsening the agreement between the simulated light curves and the observations.

The results obtained here provide further insights into the energetics and dynamics of transient events occurring in the solar corona. Further work is required, possibly with multi-stranded simulations, with more high resolution observations to study their roles in the coronal heating. One possible direction is to study the "campfire" events detected by the Extreme Ultraviolet Imager \citep[EUI;][]{eui} on board the Solar Orbiter mission \citep{so}.

\acknowledgements
 We thank the referee for the useful comments and suggestions. This work is partly supported by the Max-Planck Partner Group on "Coupling and Dynamics of the Solar Atmosphere" of MPS at IUCAA. AR acknowledges financial support from University Grants Commission in form of SRF. VLK acknowledges support from NASA Contract NAS8-03060 to the Chandra X-ray Center, and the hospitality of IUCAA during several visits. We thank Srividya Subramanian for assistance with data extraction and comparison. We thank Prof. Gulab Dewangan for useful discussions. AIA and Hi-C teams are gratefully acknowledged for providing the data.  

\software{IDL~(\url{https://www.l3harrisgeospatial.com/Software-Technology/IDL}),
SolarSoft~\citep{2012ascl.soft08013F}, 
PINTofALE~\citep{2000BASI...28..475K} and 
\ebtel~\citep{Klimchuk2008,cargill}}

\bibliography{hic}
 
\end{document}